\documentclass[twocolumn,longbibliography,amsmath,amssymb,superscriptaddress]{revtex4-1}
\usepackage{amsmath} % need for subequations
\usepackage {units}
\usepackage{here}
\usepackage{graphicx}   % need for figures
\usepackage{verbatim} % useful for program listings
\usepackage{color} % use if color is used in text
\usepackage{bm}
\usepackage{subfigure} % use for side-by-side figures
\usepackage{hyperref} % use for hypertext links, including those to external documents and URLs
\usepackage[normalem]{ulem} % for correction only, to delete in the final version
\usepackage{siunitx} % for clean units

\begin{document}
\title{Edge Current and Pairing Order Transition in Chiral Bacterial Vortex}
\author{Kazusa Beppu}\email{kazu.beppu@phys.kyushu-u.ac.jp}\affiliation{Department of Physics, Kyushu University, Motooka 744, Fukuoka 819-0395, Japan}
\author{Ziane Izri}\email{ziane.izri@email.phys.kyushu-u.ac.jp}
\affiliation{Department of Physics, Kyushu University, Motooka 744, Fukuoka 819-0395, Japan}
\author{Tasuku Sato}\affiliation{Department of Mechanical Engineering, Kyushu University, Motooka 744, Fukuoka 819-0395, Japan}\author{Yoko Yamanishi}
\affiliation{Department of Mechanical Engineering, Kyushu University, Motooka 744, Fukuoka 819-0395, Japan}
\author{Yutaka Sumino}
\affiliation{Department of Applied Physics, W-FST, I$^{2}$ Plus, and DCIS, Tokyo University of Science, Niijuku 6-3-1, Tokyo 125-8585, Japan}
\author{Yusuke T. Maeda}
\affiliation{Department of Physics, Kyushu University, Motooka 744, Fukuoka 819-0395, Japan}
\date{\today}

\begin{abstract}
We report the selective stabilization of chiral rotational direction of bacterial vortices, from turbulent bacterial suspension, in achiral circular microwells sealed by an oil-water interface. This broken-symmetry, originating from the intrinsic chirality of bacterial swimming near hydrodynamically different top and bottom surfaces, generates a chiral edge current of bacteria at lateral boundary and grows stronger as bacterial density increases. We demonstrate that chiral edge current favors co-rotational configurations of interacting vortices, enhancing their ordering. The interplay between the intrinsic chirality of bacteria and the geometric properties of the boundary is a key-feature for the pairing order transition of active turbulence.
\end{abstract}
\maketitle

Turbulent flows offer a rich variety of structures at large length scales and are usually obtained by driving flows out of equilibrium\cite{zhang} while overcoming viscous dampening. A peculiar class of out-of-equilibrium fluids, stimulated from the lower scales, also present turbulence-like structures called active turbulence \cite{ramaswamy,marchetti}. For example, a dense bacterial suspension is driven out of equilibrium by the autonomous motion of self-propelled bacteria suspended therein\cite{yeomans,frey,kokot,Li}. The alignment between bacteria shapes the active turbulence patterns of collective swimming into vortices of similar size\cite{wioland1,wioland2,beppu,hamby,clement,nishiguchi2}. However, this vortical order decays over distance, making it a long-standing issue for the development of ordered dynamics at larger scales. Hence, a growing attention is paid to novel strategies to control active turbulence with simple geometric design.

\begin{figure}[tb]
\begin{center}
\includegraphics[scale=0.56,bb=0 0 452 332]{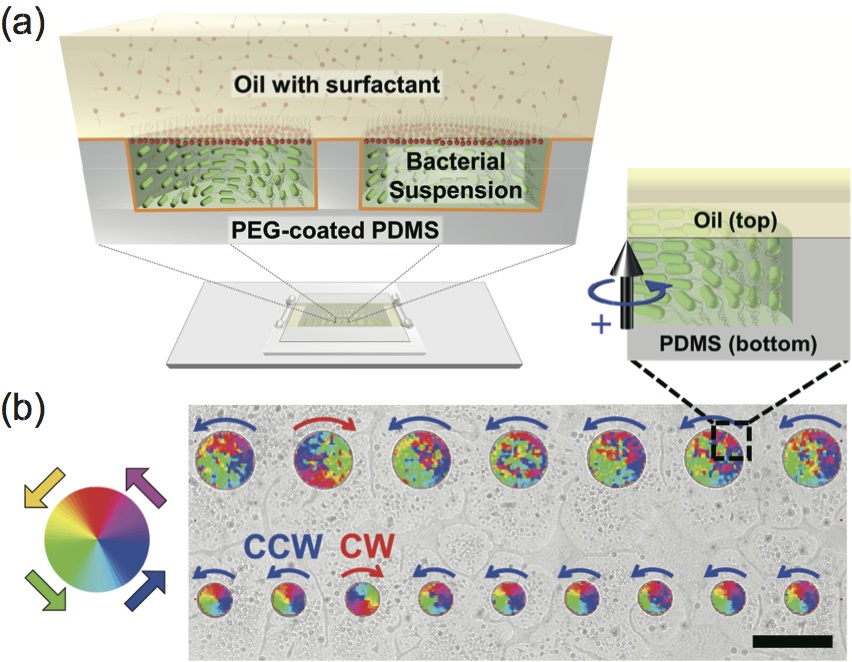}
\end{center}
\caption{\textbf{Chiral bacterial vortex.} (a) Experimental setup: a dense bacterial suspension confined in hydrophilic-treated PDMS microwells and sealed under oil/water interface stabilized with a surfactant. (b) Ensemble of chiral bacterial vortices, in microwells with radius \SI{35}{\micro\meter} (top row) and \SI{20}{\micro\meter} (bottom row). Color map codes for the direction of the velocity field. Scale bar, \SI{100}{\micro\meter}. Schematic illustration of a bacterial vortex in a single microwell. ``+'' defines the positive sign of CCW rotation. CCW and CW occurrences are displayed in blue and red arrows, respectively.}\label{fig1}
\end{figure}

Chirality, i.e. the non-equivalence of opposite handedness, is ubiquitous across scales\cite{bahr}, and is commonly involved in active systems\cite{glotzer,lowen,levis,lenz}, either biological, such as bacteria\cite{whitesides,haoran,howard,petroff}, cytoskeletons and molecular motors\cite{tee,frey2,kim}, or non-biological, consisting of self-propelled colloids\cite{jiang,bechinger1,irvine}. One of the effects of chirality is the non-equivalence of clockwise (CW) and counter-clockwise (CCW) rotations. As for bacteria, broken mirror-symmetry in flagellar rotation (CCW rotation around the tail-to-head direction during swimming) results in the opposite rotation of the cell body, which generates a net torque onto the solid surface the bacterium swims over, and in turn bends its trajectory circularly\cite{whitesides}. Despite such intrinsic chirality in individual motion, active turbulence reported in the past showed CW and CCW global rotational directions have equal probability, indicating that mirror symmetry was recovered at the collective level\cite{wioland2,beppu,nishiguchi2,hamby}. Can microscopic chirality of bacterial motion be transferred into the macroscopic order of collective swimming? Such question is a great challenge that would provide both fundamental understanding of active turbulence and technical applications for controlled material transport\cite{fabrizio,aronson}.

In this Letter, we report the chiral collective swimming of a dense bacterial suspension confined in an asymmetric (different top and bottom interfaces) but achiral (perfectly circular lateral interface) hydrodynamic boundary. Non-equivalence between CW and CCW collective swimming is enhanced as the bacterial density increases, and the selected CCW rotation with respect to the bottom-top direction, later referred to as ``top view'', induces persistent edge current mostly observed on bacteria swimming near the lateral boundaries. Such edge current can alter the geometric constraints ruling the self-organization of bacterial vortices by suppressing the anti-rotational mode. The extended geometric rule, which is in excellent agreement with experiment, brings new understanding of chiral active matter in order to organize larger scale flow.

\begin{figure}[t]
\begin{center}
\includegraphics[scale=0.49,bb=0 0 499 220]{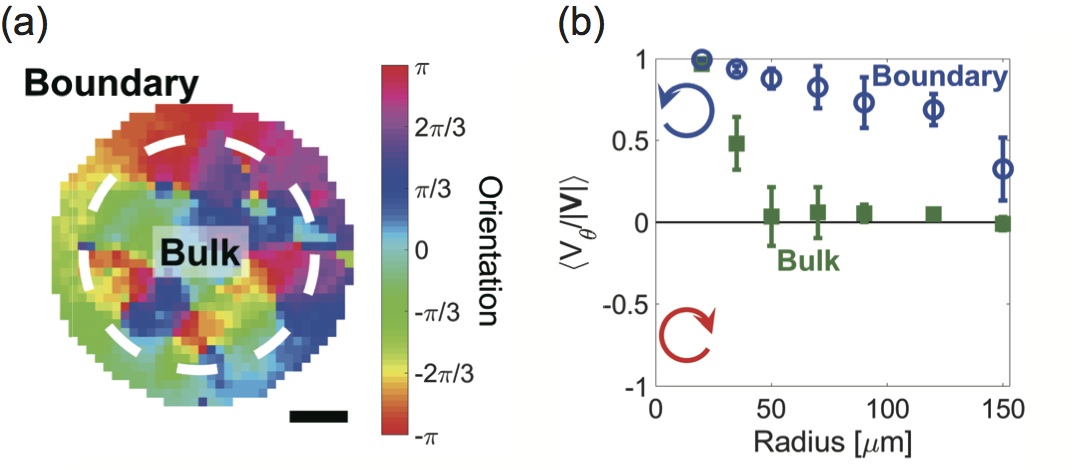}
\end{center}
\caption{\textbf{Chiral edge current.} (a) Color map of the orientation of collective motion in a chiral bacterial vortex ($R = \SI{50}{\micro\meter}$). The edge, defined as the area within \SI{10}{\micro\meter} from the lateral boundary, is separated from the rest of the microwell (the bulk) by a dashed white line. Scale bar, \SI{20}{\micro\meter}. (b) Normalized azimuthal velocity $v_{\theta}$ in microwells of various sizes. The blue circles indicate the edge current, and the green squares indicate the motion in bulk. $v_{\theta}$ is averaged over 10 s and plotted with error bars representing standard deviation.}\label{fig2}
\end{figure}

A bacterial suspension of \textit{Escherichia coli} was confined in microwells with the depth of \SI{20}{\micro\meter}, made of poly-dimethyl siloxane (PDMS) rendered hydrophilic with a polyethylene glycol coating, and sealed with an oil/water interface stabilized with a surfactant (Fig. \ref{fig1}(a))\cite{izri}. Top/bottom hydrodynamic asymmetry (later referred to as ``asymmetric conditions'') consists in the solid bottom interface of the microwell and its top fluidic interface (Fig. S1 in \cite{supplement}). Bacteria in the suspension collectively move following the circular boundary of the microwell, but show a surprisingly selective CCW rotational direction (top view). The orientation map $\theta_i$ of the velocity field $\bm{v}(\bm{r}_i)$ obtained from particle image velocimetry (PIV) shows vortical structure maintained across the microwell of $R = \SI{20}{\micro\meter}$ ( Fig. \ref{fig1}(b)) . CCW-biased vortex (called ``chiral bacterial vortex'' hereafter) occurs without built-in chirality of the confinement, e.g. ratchets\cite{fabrizio,aronson,leonardo}. CCW rotation is strongly favored at 95\% ($N = 145$, Fig. S3) probability in chiral bacterial vortex, while bacterial vortex rotates at equal probability in CCW or CW direction in water-in-oil droplets between \textit{symmetric solid (top)/solid (bottom)} interfaces (symmetric conditions, Fig. S2). Flow reversal in vortex was not observed during our observations (Fig. S3) which emphasizes that chiral bacterial vortex is much more stable than the bacterial vortices in droplets\cite{hamby}(Fig. S3). Chiral vortical structure is also observed in larger microwells ($R = \SI{35}{\micro\meter}$) but only within a distance of \SI{10}{\micro\meter} away from the circular boundary. This chiral motion with a coherent orientation near the lateral boundary, called edge current, is known to be a key feature in chiral many-body systems\cite{jiang,bechinger1,irvine}. We in turn examined the size-dependence of chiral vortices and the edge current in order to investigate the mechanism of such stability and selectivity in rotational direction. We define the tangential vector $\bm{t}(\theta_i)$ in CCW direction along the circular boundary and the azimuthal velocity $v_{\theta}(\bm{r}_i)=\bm{v}(\bm{r}_i)\cdot \bm{t}_i$. The orientation of the edge current along the boundary wall is analyzed by $\langle v_{\theta}(\bm{r}_i)/ |\bm{v}| \rangle$ where $\langle \cdot \rangle$ denotes the average over all possible site $i$ (Fig. \ref{fig2}(a)). Surprisingly, this edge current was maintained even in very large microwells ($R\geq$\SI{100}{\micro\meter}), the size of which is much larger than the critical size of a stable bacterial vortex in the bulk ($\approx\SI{35}{\micro\meter}$) (Fig. \ref{fig2}(b)). 

\begin{figure}[t]
\begin{center}
\includegraphics[scale=0.48,bb=0 0 682 510]{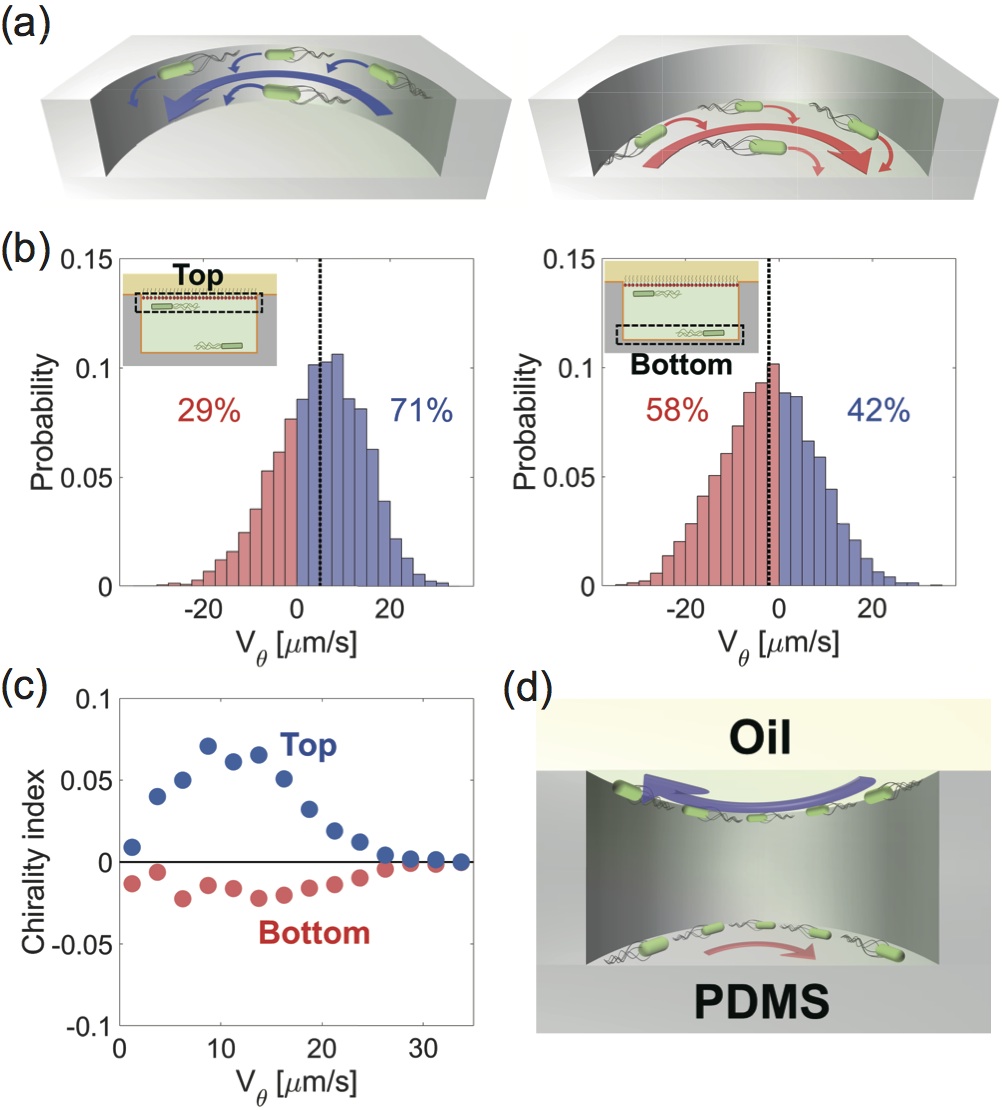}
\end{center}
\caption{\textbf{Bacterial swimming in dilute ($0.2\% v/v$) suspension} (a) Schematic illustrations of chiral bacterial swimming near the top fluidic interface (left) and near the bottom solid interface (right). Near the top interface, individual bacterial swimming is CCW-biased (blue curved arrow) while near the bottom interface it is CW-biased (see Fig. S5). (b) Histograms of azimuthal velocity and their average (vertical dashed line) of single bacteria swimming near top interface (left, schematic illustration in insert, $N=5004$, average \SI{5.02}{\micro\meter\per\second}) and bottom interface (right, schematic illustration in insert, $N=3702$, average \SI{-2.15}{\micro\meter\per\second}). Proportions of CCW and CW occurrences are displayed in blue and red at the top of each plot. (c) Corresponding chirality indices plotted against azimuthal velocity, near the top (blue) and bottom (red) interfaces. (d) Chiral bias of swimming near the top and bottom interfaces have opposite signs but different amplitudes, under asymmetric conditions.}\label{fig3}
\end{figure}

This persistence of edge current motivates us to further investigate its physical origin, by analyzing the interplay between the boundary and the intrinsic chirality of bacteria, which is the only element having chirality in the present system. With respect to the tail-to-head direction, the flagella of the bacterium rotate in the CCW direction. Torque balance then imposes a CW rotation of the body. Those two opposite rotations result in opposite frictions against the bottom interface, which ultimately converts into a CCW rotation (top view) of bacteria swimming near the top interface (CW rotation near bottom interface) (Fig. \ref{fig3}(a))\cite{whitesides}. Because of this CCW bias of their trajectories beneath the top, bacteria that collide with the lateral boundaries align with it and swim in a CCW rotation direction (CW rotation direction on the bottom). However, this effect alone does not determine net chirality because bacteria swim in opposite directions on the top and bottom interfaces of the microwell. In order to find the origin of net chirality in edge current, we recorded the trajectories of individual bacteria in dilute ($0.2\% v/v$) suspensions, such that interactions between bacteria do not affect their swimming. Fig. \ref{fig3}(b) presents the probability distribution function of the azimuthal velocities of individual bacteria, $P(v_{\theta})$, in a microwell ($R = \SI{35}{\micro\meter}$). Individual bacterial motion beneath the top fluidic interface shows visible nonequivalence between CW and CCW swimming, as more than half ($71\%$) of the tracked bacteria swim in CCW direction. By contrast, $58\%$ of individual bacteria onto the bottom solid interface swim in CW rotational direction, indicating weaker chiral bias. To compare those swimming chiralities, we define the chirality index $CI(v_{\theta})$ as antisymmetric part of $P(v_{\theta})$, i.e. 
\begin{equation}
CI(v_{\theta}) = P(v_{\theta}) - P(-v_{\theta}).
\end{equation}
Positive chirality index means CCW rotation is dominant, and the larger the absolute value of the index, the more biased the rotational direction. For bacteria swimming near the top interface, chirality index is positive, while it is negative near the bottom interface (Fig. \ref{fig3}(c)). This tells that bacterial motion is CCW-biased near the top interface, and CW near the bottom interface. In addition, near the top interface, $|CI(v_{\theta})|>0.05$ whereas near the bottom interface $|CI(v_{\theta})| < 0.02$.  This indicates that bacterial swimming is more biased near the top (fluidic) interface than near the bottom (solid) interface. This difference in the amplitude of the bias at each interface is responsible for the chiral rotation of the whole bacterial vortex, and therefore is at the origin of chiral edge current (Fig. \ref{fig3}(d)). 

\begin{figure}[t]
\begin{center}
\includegraphics[scale=0.57,bb=0 0 440 394]{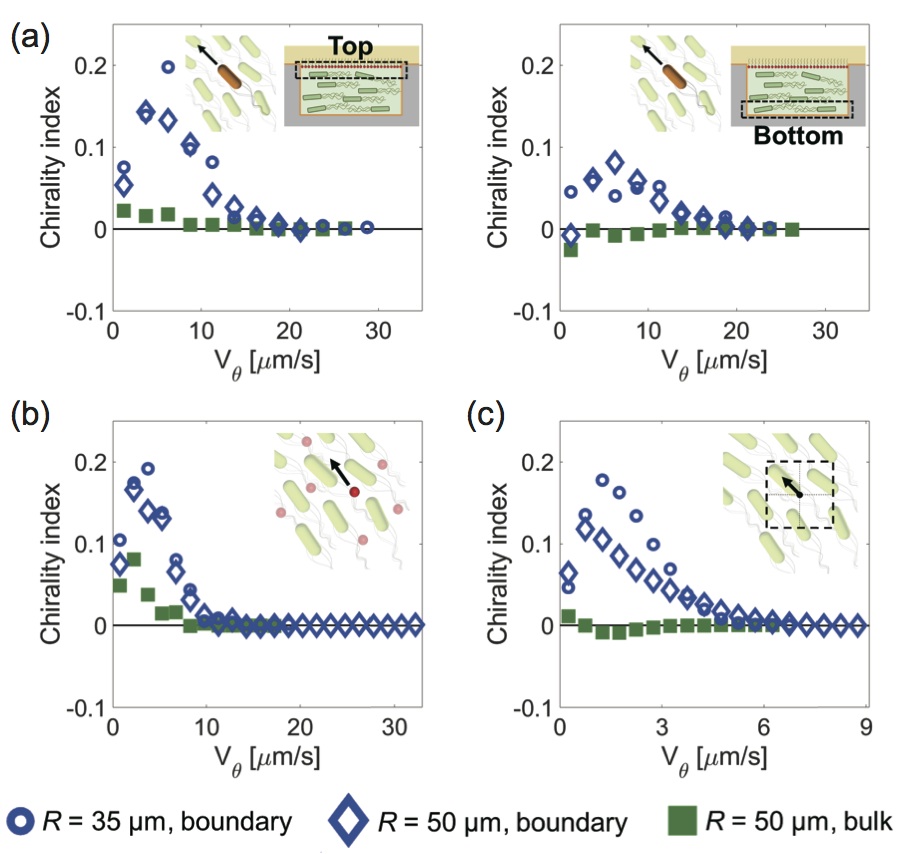}
\end{center}
\caption{\textbf{Chiral bacterial vortex is collective effect.} A dense ($20\% v/v$) bacterial suspension is confined under asymmetric conditions. Represented chirality indices against azimuthal velocity of (a) individual fluorescent bacteria (schematic illustration in insert) observed under confocal microscopy near the top (left) and bottom (right) interface, (b) fluid flow with tracer particles also observed under confocal microscopy, and (c) collective bacterial motion observed under bright field. Two microwell sizes were considered: $R = \SI{35}{\micro\meter}$ (small blue circles) and $R = \SI{50}{\micro\meter}$. In the larger microwells were considered two regions: the boundary layer within \SI{10}{\micro\meter} from the lateral boundary (larger blue diamonds), and the bulk that is more than \SI{10}{\micro\meter} away from the lateral wall (green disks). Sample size and average azimuthal velocity of each case can be found in Fig. S6.}\label{fig4}
\end{figure}
As the density of self-propelled particles increases, their mutual interactions affect more significantly their global motion, which ultimately gives rise to collective behavior. As implied already, the rotational biases of handedness of individual bacteria ($71\%$ in top and $58\%$ in bottom) are too weak to account for the 95\%-selective chirality in the bacterial vortex (Fig. \ref{fig1}(c)). To resolve this gap, we therefore characterized the structure of chiral bacterial vortices in much denser ($20\% v/v$) suspensions. Fig. \ref{fig4}(a) presents chirality index of individual bacteria near the top and bottom interfaces in a microwell ($R = \SI{35}{\micro\meter}$). Interestingly, both interfaces present dominantly CCW rotational direction, although the top interface is more strongly biased. This indicates that the effect of interactions between bacteria forces the rotational direction to be the same across the microwell. Intriguingly, chirality index of individual motion becomes larger as the bacterial density increases. Trajectories of tracer particles were also predominantly CCW, indicating that fluid flow also has CCW handedness. Moreover, chirality index of the fluid velocity (Fig. \ref{fig4}(b)) and the collective velocity (Fig. \ref{fig4}(c)) are comparable to that of individual motion. This suggests that a slight chiral bias in individual bacteria is amplified by collective bacterial interactions, which turns into a global vortex with a unidirectional rotation.

When we analyzed the vortex flow near the lateral boundary of larger microwells ($R = \SI{50}{\micro\meter}$), individual motion at high density has a large positive chirality index, similarly to smaller microwells. However, near the center of larger microwells, chirality is null for individual and collective bacterial motion as well as for the fluid flow (Fig. \ref{fig4}(a) to (c)). The necessity to be close to the lateral boundaries to observe coherent chiral swimming emphasizes the importance of spatial constraint to amplify the rotational bias and turn it into chiral edge current.

\begin{figure}[tb]
\begin{center}
\includegraphics[scale=0.55,bb=0 0 556 537]{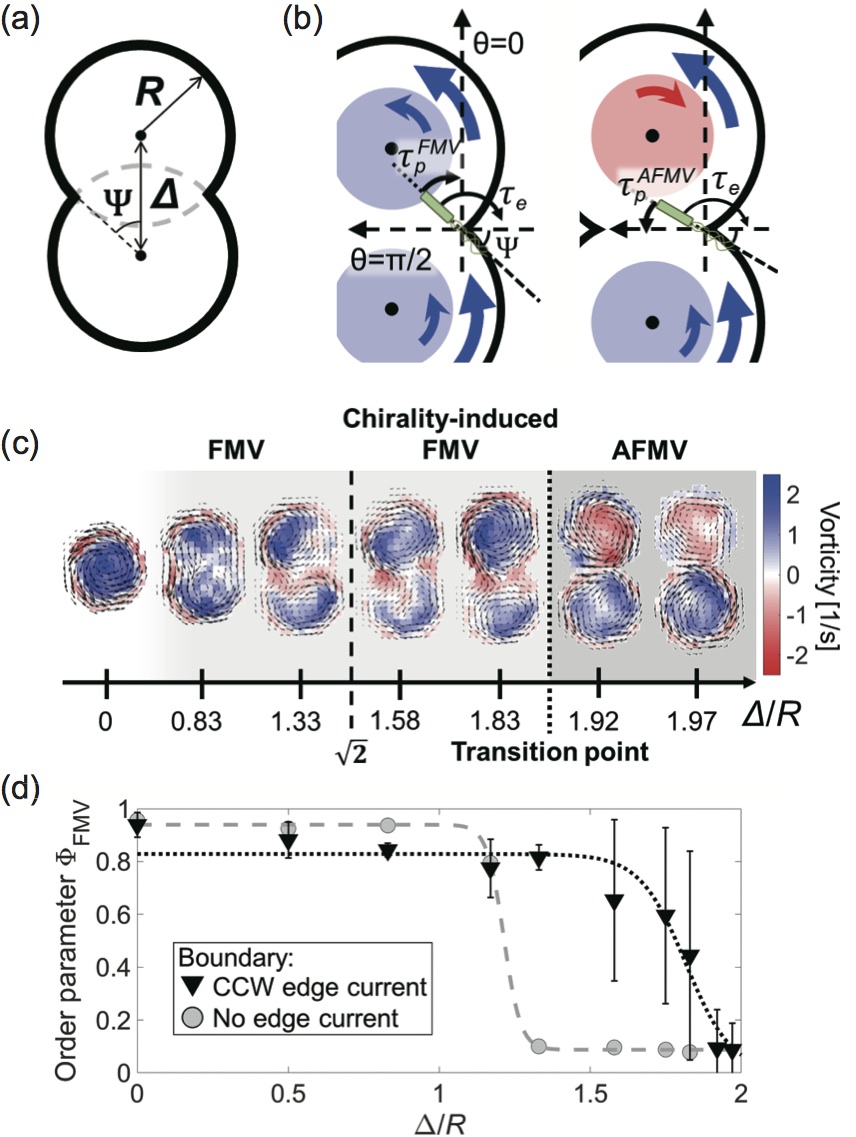}
\end{center}
\caption{\textbf{Edge current favors co-rotational vortex pairing} (a) Schematic illustration and definition of relevant geometric parameters. (b) Illustration of the co-rotational vortex pairing (FMV pattern, left) and the anti-rotational pairing (AFMV pattern, right) with edge current. The edge current deviates the orientation angle of bacteria $\theta$ around the \textit{tip}. In FMV pattern, bi-particle alignment and edge current deviate bacteria in the same direction, but in AFMV pattern, those effects compete. (c) Vorticity map of bacterial vortex pairs at various values of $\Delta$ with $R = \SI{19}{\micro\meter}$. (d) Order parameter $\Phi_{FMV}$ of FMV pairs of interacting bacterial vortices, against $\Delta/R$. Under no edge current (full grey circles), FMV to AFMV transition occurs at $\Delta/R\simeq\sqrt{2}$, while under CCW edge current (inverted black triangles) it occurs at a larger value of $\Delta/R\simeq1.9$.}\label{fig5}
\end{figure}

The edge current plays a crucial role in the ordering of interacting bacterial vortices. When two bacterial vortices interact with one another via near-field interaction, they have either the same rotational direction (defined as ferromagnetic vortices, FMV) or opposite rotational directions (anti-ferromagnetic vortices, AFMV) in a geometry-dependent manner\cite{wioland2,beppu,nishiguchi2}. To reveal how a chiral edge current affects the ordering of interacting bacterial vortices, we construct a theoretical model of interacting bacterial vortices in doublets of overlapping identical circular boundaries (Fig. \ref{fig5}(a))\cite{beppu}. 

Two identical overlapping circular microwells, with a radius $R$ and an inter-center $\Delta$, offer the means for a systematic investigation of the pairing order transition\cite{beppu}. The ratio $\Delta/R = 2\cos\Psi$ is an important geometric parameter characterizing the pairing order transition from FMV to AFMV: if $\Delta$/R is small enough, the two vortices align in co-rotational direction and pair into a FMV pattern. Transition from FMV to AFMV occurs at a predicted value $\Delta_c/R = 2 \cos \SI{45}{\degree} = \sqrt{2}$ because that is the only configuration at which the two pairing patterns are equiprobable\cite{beppu}. How does edge current affect the previously established design principle?

To answer this question, the orientation dynamics of bacteria with the heading angle $\theta$ is considered at the vicinity of the sharp areas of the boundaries (``\textit{tip}''). Given the CCW fluid flow near boundary reorients the bacteria at the \textit{tip}, the effective torque for reorientation $\bm{\tau_e} = - \partial U_e/\partial \theta$ has geometry-dependent potential $U_e = -2 \gamma_e \sin\theta\cos\Psi$ with $\gamma_e$ representing the ratio of fluid flow to bacterial swimming ($\gamma_e > 0$). This reorientation maintains the CCW rotation of the edge current along boundary in both FMV pairing (Fig. \ref{fig5}(b), left) and AFMV pairing (Fig. \ref{fig5}(b), right). In addition, the result of the bi-particular collision near the \textit{tip} between bacteria coming from different circular parts\cite{beppu} also affects the vortex pairing. Bacterial collision is ruled by a polar alignment as a source of geometry-dependence\cite{vicsek,supplement}. By considering the most probable configurations (general case detailed in \cite{supplement}), the orientation near the \textit{tip} is decided by the potential of FMV pairing $U_p^{FMV} = 2\gamma_p\sin\Psi$ at $\theta = 0$ (Fig. \ref{fig5}(b), left) or of AFMV pairing $U_p^{AFMV} = 2\gamma_p\cos\Psi$ at $\theta = \pi/2$ (Fig. \ref{fig5}(b), right), with the strength of the polar alignment $\gamma_p>0$. The respective sums of the potentials coincide at the transition point, i.e. $U_p^{FMV} + U_e|_{\theta=0} = U_p^{AFMV} + U_e|_{\theta=\pi/2}$, which leads to
\begin{equation}
\gamma_p \sin\Psi_c = (\gamma_p- \gamma_e) \cos\Psi_c,
\end{equation}
where $\gamma_p- \gamma_e$ indicates the suppression of AFMV pairing by chiral edge current. Hence, $\Delta_c/R=2\cos\Psi_c$ at which FMV and AFMV pairings occur at equal probability is
\begin{equation}\label{chiral_geometric rule}
\frac{\Delta_c}{R} = \frac{2}{\sqrt{1+(1-\gamma_e/\gamma_p)^2}}\simeq \sqrt{2} \Bigl( 1 + \frac{\gamma_e}{2\gamma_p}\Bigr).
\end{equation}
FMV pattern is stabilized in $\Delta/R \geq \sqrt{2}$ and the relative strength of the edge current $\gamma_e/\gamma_p$ determines the shift of the transition point.

To test those chirality effects, we examined the pairing of doublets of chiral vortices with $R = \SI{19}{\micro\meter}$ within the range of $0 \leq \Delta < 2R = \SI{38}{\micro\meter}$. FMV pattern of chiral vortices is dominant in $0 \leq \Delta/R \leq 1.9$ and exhibits CCW rotation (Fig. \ref{fig5}(c)). The pairing order transition is also analyzed by using the order parameter of FMV pairing $\Phi_{FMV}$ \cite{supplement}. $\Phi_{FMV}$, which reaches 1 for FMV while it goes down to 0 for AFMV, is defined as $\vert \langle \bm{p}_i \cdot \bm{u}_i \rangle \vert$ with the orientation of velocity $\bm{p}_i$ measured experimentally at site $i$ and the expected orientation of FMV pattern $\bm{u}_i$ calculated numerically at corresponding site. Under asymmetric condition, $\Phi_{FMV}$ shows a transition from $1$ to $0$ at $\Delta/R\simeq1.9$, while it occurs at $\Delta/R\simeq1.4$ under symmetric condition (absence of chiral edge current) (Fig. \ref{fig5}(d)). FMV pairing pattern appears to be favored in the presence of edge current. Moreover, according to Eq.\eqref{chiral_geometric rule} with the experimental values $\gamma_p=0.5$ (obtained from independent experiment, Fig. S7) and $\Delta_c/R\simeq1.9$, the coefficient of edge current is $\gamma_e=0.3$ that is reasonably comparable to the ratio of fluid flow to bacterial swimming velocity (Figs.\ref{fig4}(a) and 4(b)). The excellent agreement with experiment means that chiral edge current affects the pairing order transition.

In conclusion, we revealed that confining a dense bacterial suspension in microwells with asymmetric top/bottom interfaces but achiral circular lateral boundaries stabilizes a chiral vortex. Hydrodynamically different top/bottom interfaces give then rise to a subtle broken symmetry in bacterial swimming, that is amplified by collective bacterial motion and becomes an edge current persistent over larger length scale. We showed that it is possible to generate and control the break of mirror-symmetry without using built-in chirality, unlike most current experimental setup, e.g. with built-in chiral ratchet-shape\cite{fabrizio,aronson,leonardo}. Moreover, in present chiral bacterial vortex, rigid rod much longer than bacterial body can consistently rotate in CCW direction over multiple rounds at \SI{0.5}{\radian\per\second} (Fig. S8). Thus, asymmetric hydrodynamic boundary offers simple and fast material transport, even without built-in chirality.
Finally, the edge current favors co-rotational FMV pairing pattern of doublets of identical vortices. Even in triplets of identical overlapping circular microwells, the pairing order transition from FMV to (frustrated) AFMV patterns is also shifted to higher values (from $1.7$ to $1.9$) (Fig. S9), suggesting that chiral bacterial vortex has fewer limitation from geometric frustration. Although beyond the scope of this study, understanding how the nature of an interface affects the amplitude of the chiral flow it generates could give the means of a finer tuning of the global rotation of confined bacterial vortices. 

The finding of edge current in chiral bacterial vortex opens new directions for the tailoring of collective motion. Such asymmetric hydrodynamic boundaries are also involved in flocking and lane formation of active droplets\cite{stone}. The edge current observed in active droplets with similar collective effect may advance the generic understanding of chiral collective behavior\cite{glotzer,tjhung}. Furthermore, stabilized co-rotational pairing order is a key to clarify how broad class of active matter amplifies microscopic chirality. As such constrained pairing order is also relevant to chiral spinners\cite{tierno}, controlling chirality-induced order with simple geometric rule would verify the validity of geometric approach to chiral many-body systems.

This work was supported by Grant-in-Aid for Scientific Research on Innovative Areas (JP18H05427 and JP19H05403) and Grant-in-Aid for Scientific Research (B) JP17KT0025 from MEXT.

\newpage

$ $

\newpage

\onecolumngrid
\setcounter{figure}{0}
\renewcommand{\thefigure}{S\arabic{figure}}
\renewcommand{\figurename}{FIG.}
\renewcommand{\tablename}{TABLE.}
\renewcommand{\thetable}{S\arabic{table}}
\renewcommand{\refname}{References}
\renewcommand{\arraystretch}{1.2}

\section*{Supplemental information for \\Edge Current and Pairing Order Transition in Chiral Bacterial Vortex} 
\author{Kazusa Beppu}\affiliation{Department of Physics, Kyushu University, Motooka 744, Fukuoka 819-0395, Japan}
\author{Ziane Izri}\affiliation{Department of Physics, Kyushu University, Motooka 744, Fukuoka 819-0395, Japan}
\author{Tasuku Sato}\affiliation{Department of Mechanical Engineering, Kyushu University, Motooka 744, Fukuoka 819-0395, Japan}\author{Yoko Yamanishi}
\affiliation{Department of Mechanical Engineering, Kyushu University, Motooka 744, Fukuoka 819-0395, Japan}
\author{Yutaka Sumino}
\affiliation{Department of Applied Physics, W-FST, I$^{2}$ Plus, and DCIS, Tokyo University of Science, Niijuku 6-3-1, Tokyo 125-8585, Japan}
\author{Yusuke T. Maeda}
\affiliation{Department of Physics, Kyushu University, Motooka 744, Fukuoka 819-0395, Japan}

\maketitle
\section{Materials and Methods}
\subsection{Device microfabrication}

The device used in this experiment is a flow cell that contains an array of microwells of various radii $R$ ($\SI{20}{\micro\meter} \leq R \leq \SI{150}{\micro\meter}$) and three types of shape: single circular microwells, pairs and triplets of overlapping identical circular microwells. Their depth is \SI{20}{\micro\meter} in all the experiments\cite{beppu}. The flow cell is made of a cover glass slide (Matsunami, S1127, \SI{1.0}{\milli\meter}) and a cover slip (Matsunami, C218181, \SI{0.1}{\milli\meter}) separated by a double-sided adhesive tape (NICHI-BAN, NW-10, \SI{100}{\micro\meter}).

The microwells are fabricated using standard soft lithography techniques. Briefly, PDMS polymer and curing agent (Dow Corning, 98-0898, Sylgard184) at 90-to-10 mass ratio was spin-coated at 1000 rpm for 30 seconds to reach a thickness of \SI{100}{\micro\meter}, on a mold made of a photoresist (SU-8 3025, Microchem) pattern (thickness \SI{20}{\micro\meter}) cured and developed through conventional photo-lithography on a silicon wafer (Matsuzaki, Ltd., $\phi$2-inch wafer). After a curing at \SI{75}{\celsius} for an hour, the PDMS film is cut around a single array of microwells and peeled off to be bonded on its unpatterned surface to the glass cover slide that has been exposed to air plasma beforehand (\SI{10}{\second}, corona discharge gun, Shinko Denso).

\begin{figure}[b]
\begin{center}
\includegraphics[scale=0.67,bb=0 0 700 252]{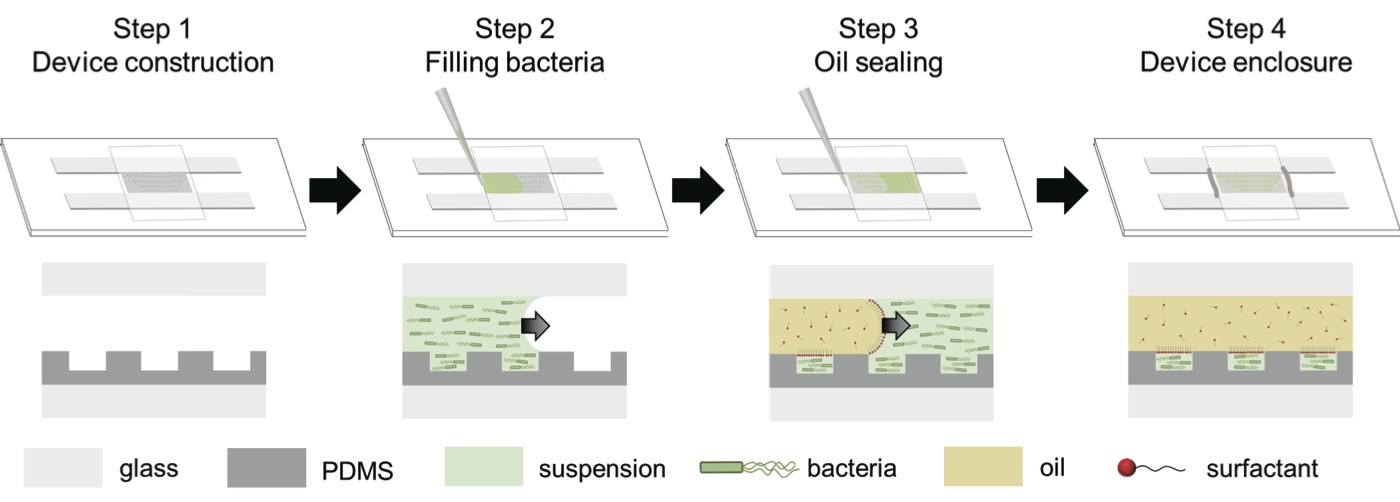}
\end{center}
\caption{\textbf{Schematic illustration of device fabrication.} Step 1: device construction. A thin PDMS layer patterned with microwells is transferred between the two glass slides of a flow cell and coated with PEG-PLL to avoid unspecific adhesion of bacteria. Step 2: bacterial suspension is injected from one side of the flow cell. Step 3: excess bacterial suspension is flushed out by a mixture of oil and surfactant. This mixture seals the microwells on their top side. Step 4: device is sealed at both ends with epoxy glue.}\label{fig.s1}
\end{figure}

This assemblage is then once again exposed to air plasma (\SI{10}{\second}, corona discharge gun), covered with a solution of polyethylene glycol-poly-L-lysine (PEG-PLL, Nanocs, PG2k-PLY), and left to rest for 30 minutes. Ungrafted PEG-PLL is then washed away with deionized water. PDMS microwells and glass cover slide are now treated hydrophilic, and PEG-PLL coating prevents non-specific adhesion of bacteria. Finally, the flow cell is completed with the glass cover slip (untreated) being attached to the glass cover slide with the adhesive spacer placed over the unpatterned areas of the PDMS sheet. It is used immediately after fabrication (Fig. \ref{fig.s1} Step 1).

\subsection{Filling protocol}
Once a flow cell is ready, it is filled with \SI{20}{\micro\liter} of a bacterial suspension (two volume fractions were available: dense at $20\% v/v$, dilute at $0.2\% v/v$). The PEG-PLL coating of the PDMS allows the bacterial suspension to reach the inside of the microwell and fill them properly. Once the flow cell is completely filled, oil (light mineral oil, Sigma Aldrich) with surfactant (SPAN80, Nacalai) at $2 wt\%$ is injected from the same side, while excess bacterial suspension over the microwells is flushed out and absorbed with filter paper from the other side of the flow cell. This seals the microwells under an oil/water interface. To suppress unwanted flow in the flow cell, both of its ends are sealed with epoxy glue (Huntsmann, Ltd.). The array of microwells is then ready to be observed under microscope (Fig. \ref{fig.s1} Steps 2-4)\cite{izri}.

\subsection{Persistently straight-swimming bacterial strain}
We used bacterial strain \textit{Escherichia coli} RP4979 that lacks tumbling ability. These bacteria swim smoothly without tumbling and show persistent straight motion in the bulk of a fluid. Highly motile bacteria were obtained by inoculating a single bacterial colony into \SI{1.5}{\milli\liter} of LB medium (NaCl \SI{5}{\gram\per\liter}, yeast extract \SI{5}{\gram\per\liter}, tryptone \SI{10}{\gram\per\liter}, pH7.2) and incubating it overnight at \SI{37}{\celsius}. The next day, \SI{50}{\micro\liter} of this overnight culture solution is transferred to \SI{25}{\milli\liter} of T-Broth (NaCl \SI{5}{\gram\per\liter}, trypton \SI{10}{\gram\per\liter}, pH7.2), and the inoculated cultures are incubated at \SI{30}{\celsius} and agitated at 150 rpm for about 6 hours. After reaching an optical density of $0.4$, the culture medium is centrifuged at 3000 rpm at room temperature for 10 minutes to concentrate the bacterial suspension density to $20-25\% v/v$.

\begin{figure}[h]
\begin{center}
\includegraphics[scale=0.57,bb=0 0 600 272]{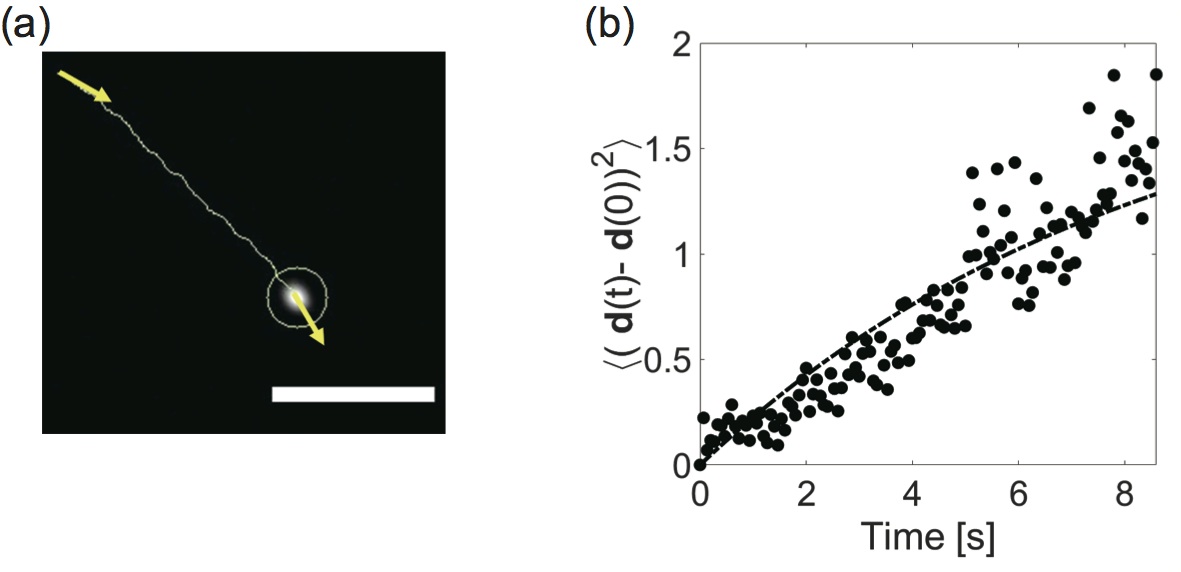}
\end{center}
\caption{\textbf{Straight swimming of single bacteria in bulk.} (a) Typical trajectory of swimming single bacteria in bulk. Scale bar is \SI{20}{\micro\meter}. (b) Mean square displacement of heading angle of single bacteria is plotted with time. The slope of this MSD curve is fitted with $2(1-\exp[-Dt])$, where $D$ reflects orientation fluctuation of smoothly swimming bacteria\cite{jain}.}\label{figs_diffusion}
\end{figure}

The straight swimming of bacteria was measured in a bulk fluid, and the fluctuation of the heading angle $\theta(t)$ was analyzed. Single bacteria are considered as self-propelled points particle, with a position $\bm{r}(t) = (\bm{x}(t), \bm{y}(t))$ and an orientation $\bm{d}(t)$(Fig.\ref{figs_diffusion}(a)). For two-dimensional coordinates, the orientation of single bacteria is expressed by the unit vector along the long-axis of bacteria $\bm{d} = \bm{d}(\theta(t)) = (\cos\theta(t) , \sin\theta(t))$. Because the mean-square angle displacement (MSD) is given by $\langle \bigl[\bm{d}(\theta(t)) - \bm{d}(\theta(0))\bigr]^2 \rangle_t = 2(1-\exp[-D t])$ with time interval $\delta t$ and the angular diffusion coefficient $D$ that reflects the fluctuation of bacterial orientation at single cell level\cite{jain}(Fig.\ref{figs_diffusion}(b)). The obtained coefficient $D$ is 0.12 rad$^2$/sec, indicating that RP4979 bacteria persistently swim in one direction at a low density in bulk. 

\subsection{Bacterial density measurement and image velocimetry}

To measure bacterial density, we used a mixture - 99-to-1 ratio - of two genetically modified bacteria (strain RP4979) that constitutively express fluorescent protein (either YFP or dTomato). That fluorescent labeling allows the quantitative recording of the trajectories of individual bacterial swimming in either dilute (Fig. 3 in main text) or dense suspensions (Fig. 4(a) in main text). By tracking dTomato-expressing bacteria, we can record the individual trajectories of bacteria inside collective vortical motion.

In the analysis of single bacteria and tracer particles in a suspension, they were tracked by means of a plugin of Particle Tracker 2D/3D in the Fiji(ImageJ) software. Bright-field optical imaging and video-microscopy were performed by using an inverted microscope (IX73, Olympus) with a CCD camera (DMK23G445, Imaging Source) that enables us to record bacterial collective motion at 30 frames per second. The velocity field of bacterial collective motion $\bm{v}(r,t)$ was analyzed by Particle Image Velocimetry(PIV) with Wiener filter method using PIVlab based on MATLAB software, and its grid size was \SI{2.98}{\micro\meter} $\times$\SI{2.98}{\micro\meter}. Acquired velocity fields were further smoothed by averaging over 1 sec.

In addition, \SI{0.5}{\micro\meter} polystyrene tracer particles with red fluorescence (Molecular probes) were used to track the flow field (Fig. 4(b) in main text). The tracer particles were dispersed at a low density of $0.026\% v/v$, where individual particles could be tracked inside the bacterial suspension. Recording of the trajectories of red-labeled bacteria and red tracer particles was done with a confocal microscope (IX73 inverted microscope from Olympus, and confocal scanning unit CSU-X1 from Yokogawa Electric Cor. Ltd., iXon-Ultra EM-CCD camera from Andor Technologies) under red fluorescence channel. All the recordings were done at 30 frames per second.

\section{Experimental details}
\subsection{Preponderance and stability of CCW bacterial vortex}

\begin{figure}[b]
\begin{center}
\includegraphics[scale=0.67,bb=0 0 400 452]{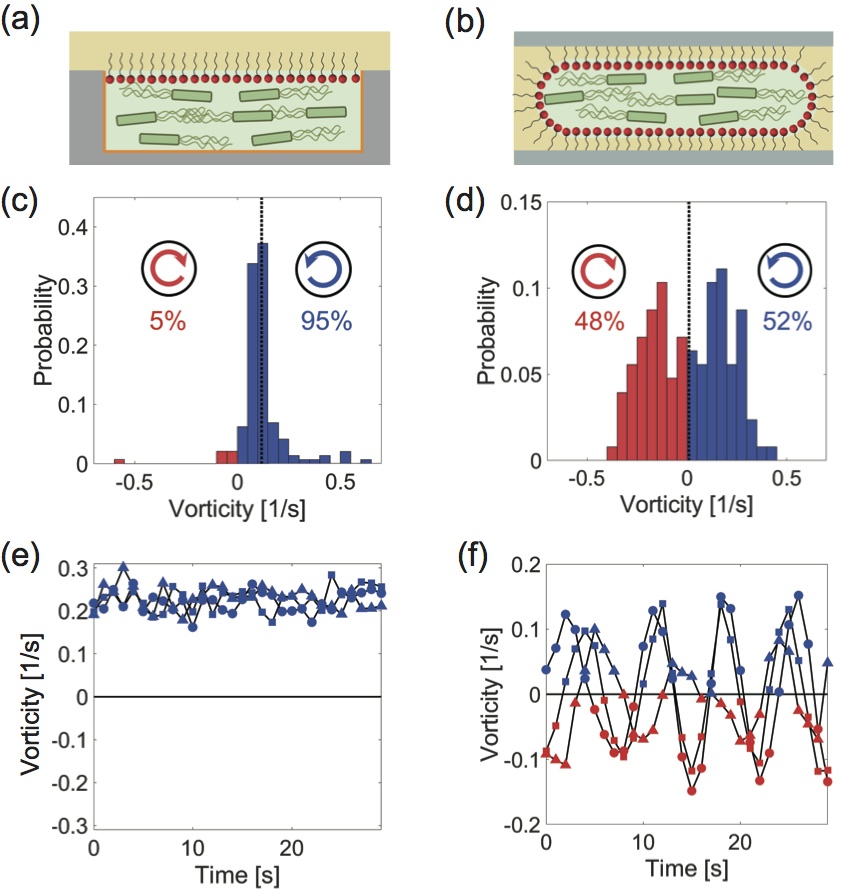}
\end{center}
\caption{\textbf{Preponderance and stability of chiral bacterial vortex.} (a) A droplet of a dense bacterial suspension confined in a microwell ($R = \SI{20}{\micro\meter}$) between \textit{asymmetric fluidic (top) and solid (bottom)} interfaces. (b) Control experiment for the effect of top/bottom interfaces on the directionality of the bacterial vortex. A droplets ($\SI{15}{\micro\meter}\leq R\leq\SI{35}{\micro\meter}$) of an emulsion of dense bacterial suspension in oil between \textit{symmetric solid (top)/solid (bottom)} interfaces. (c) Distributions of vorticity averaged over 10 s and their respective average (vertical dashed line) under asymmetric interfaces ($N = 145$, average \SI{0.12}{\per\sec}) in a microwell ($R = \SI{20}{\micro\meter}$). Proportions of CCW and CW occurrences are displayed in blue and red at the top of each plot. (d) The distributions of vorticity and their respective average (vertical dashed line) under symmetric solid/solid interfaces ($N = 126$, average \SI{0.01}{\per\sec}). Under symmetric interfaces, CW and CCW rotations are observed with the same frequency. (e) Persistent dynamics of vorticity. The vorticity is always positive without sign change, meaning CCW rotation is stable over the range of our measurement (30 sec). (f) Dynamics of vorticity change under solid / solid symmetric interfaces show the frequent change of vorticity sign, indicating that switching between CW and CCW rotations occurs.}\label{fig.s2}
\end{figure}

The chiral bias of bacterial swimming can be affected by the nature of the top and bottom interfaces, i.e. whether the two interfaces are hydrodynamically equivalent or not as seen in main text. We compare this effect on chiral bacterial vortices (dense suspension) between asymmetric (Fig. \ref{fig.s2}(a)) and symmetric top/bottom interfaces (Fig. \ref{fig.s2}(b)). Fig. \ref{fig.s2} (c) shows the distribution of vorticity of an ensemble of chiral bacterial vortices. CCW rotation is favored between \textit{asymmetric fluidic (top)/solid (bottom)} interfaces (later referred to as asymmetric conditions) and the probability of CCW rotation is 95\%. We next tested the vortex rotation of bacterial collective motion under symmetric top/bottom interface. We confined dense bacterial suspension in water-in-oil droplets between \textit{symmetric solid (top)/solid (bottom)} interfaces (symmetric conditions). By examining the vorticity of bacterial collective motion an emulsion of $\SI{15}{\micro\meter}\leq R\leq\SI{35}{\micro\meter}$, a vortex rotated at equal probability in CCW (52\%) or CW direction (48\%) (Fig. \ref{fig.s2}(d)). Thus, the selective chirality is observed only in the asymmetric top/bottom interfaces.
Moreover, such prepondence in chirality can be involved in the stability of rotational direction of vortical flow in dynamics. we analyzed the dynamics of vorticity for 30 sec in order to examine the flow reversal in chiral bacterial vortex. For chiral bacterial vortex under asymmetric condition, the sign reversal was not observed in this typical observation time (Fig. \ref{fig.s2}(e)). In contrast, for achiral bacterial vortex under symmetric condition, the vorticity shows periodical change with sign reversal as reported in previous study\cite{hamby} (Fig. \ref{fig.s2}(f)). Thus, chiral bacterial vortex is highly selective (95\% in CCW rotation) and stable (without flow reversal for few tens of sec).

\subsection{Edge current is not observed in a microwell with symmetric top/bottom condition}
\begin{figure}[hb]
\begin{center}
\includegraphics[scale=0.6,bb=0 0 600 252]{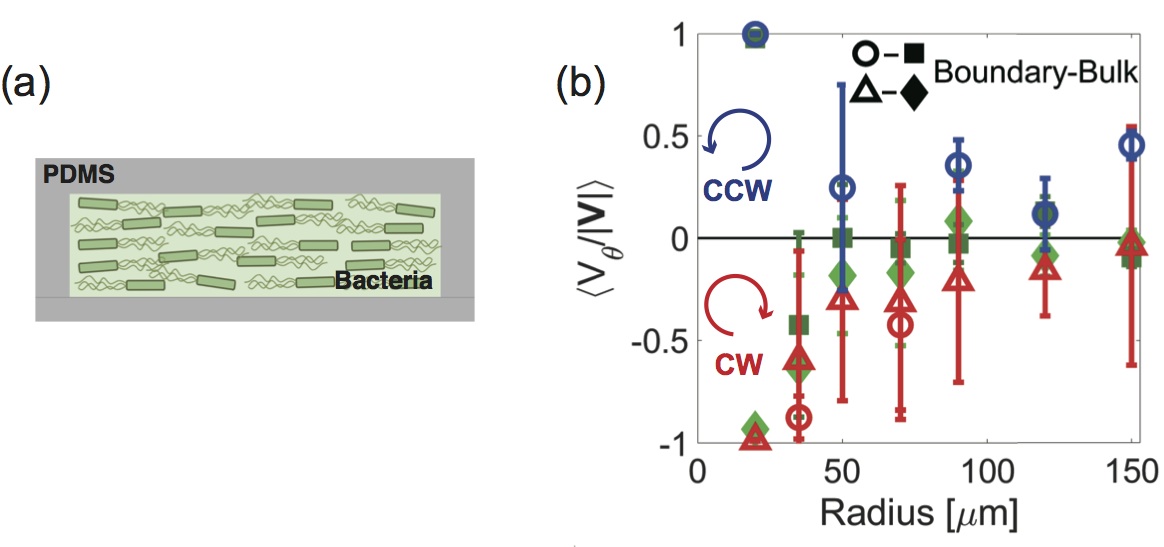}
\end{center}
\caption{\textbf{Weak edge current of bacteria in a microwell with symmetric condition.} (a) Schematic illustration of a dense bacterial suspension confined in a microwell between \textit{symmetric solid (top)/solid (bottom)} interfaces. (b) The edge current in symmetric PDMS microwell. The edge, defined as the area within \SI{10}{\micro\meter} from the lateral boundary, is separated from the rest of the microwell (the bulk) as same as Figure 2 in main text. Normalized azimuthal velocity of the flow in microwells of various sizes. The blue circles (red circles) indicate the edge current in CCW direction (in CW direction), and the green squares indicate the vortical collective motion in bulk. Normalized azimuthal velocities averaged over 10 s are plotted with error bars that present standard deviations of their time series.}\label{fig.edge}
\end{figure}

In Figure 2 in main text, we showed the edge current in CCW rotation in chiral bacterial vortex. This edge current is also observed in larger microwells but only within a distance of \SI{10}{\micro\meter} away from the circular boundary. To test whether this edge current is unique to the spatial confinement between \textit{asymmetric fluidic (top)/solid (bottom)} interfaces, we also examined the bacterial vortex in the circular microwell between \textit{symmetric solid (top)/solid (bottom)} interfaces (Fig. \ref{fig.edge}(a)). 

The solid interface was made of PEG-coated PDMS and the dense bacterial suspension (20\% volume fraction) was confined in microwells. We define the tangential vector $\bm{t}(\theta_i)$ in CCW direction along the circular boundary and the azimuthal velocity $v_{\theta}(\bm{r}_i) = \bm{v}(\bm{r}_i)\cdot \bm{t}_i$. The edge, defined as the area within \SI{10}{\micro\meter} from the lateral boundary, is separated from the rest of the microwell (the bulk). The orientation of the edge current along the boundary wall is analyzed by $\langle v_{\theta}(\bm{r}_i)/ |\bm{v}| \rangle$  where $\langle \cdot \rangle$ denotes the average over all possible site $i$ and observation time of 10 sec (Fig. 2(a) in main text). However, the stable edge current in symmetric conditions is observed only in smaller microwells with $R\leq$\SI{35}{\micro\meter}(red open triangle: CW direction, blue open circle: CCW direction, in Fig. \ref{fig.edge}(b)). The threshold value \SI{35}{\micro\meter} is comparable to the critical size of a stable bacterial vortex in the bulk (green filled square in Fig. \ref{fig.edge}(b)), and this characteristic length is much shorter than the one for chiral edge current ($\approx \SI{100}{\micro\meter}$, see Fig 2(b) in main text). In addition, collective motion near boundary has no preference of either CW or CCW directions in symmetric conditions. Thus, the collective motion in symmetric conditions does not have persistent edge current in CCW rotation, indicating that chiral edge current is uniquely stabilized in asymmetric top/bottom conditions.

\subsection{Trajectory of single bacteria near solid interface}

\begin{figure}[hb]
\begin{center}
\includegraphics[scale=0.5,bb=0 0 800 602]{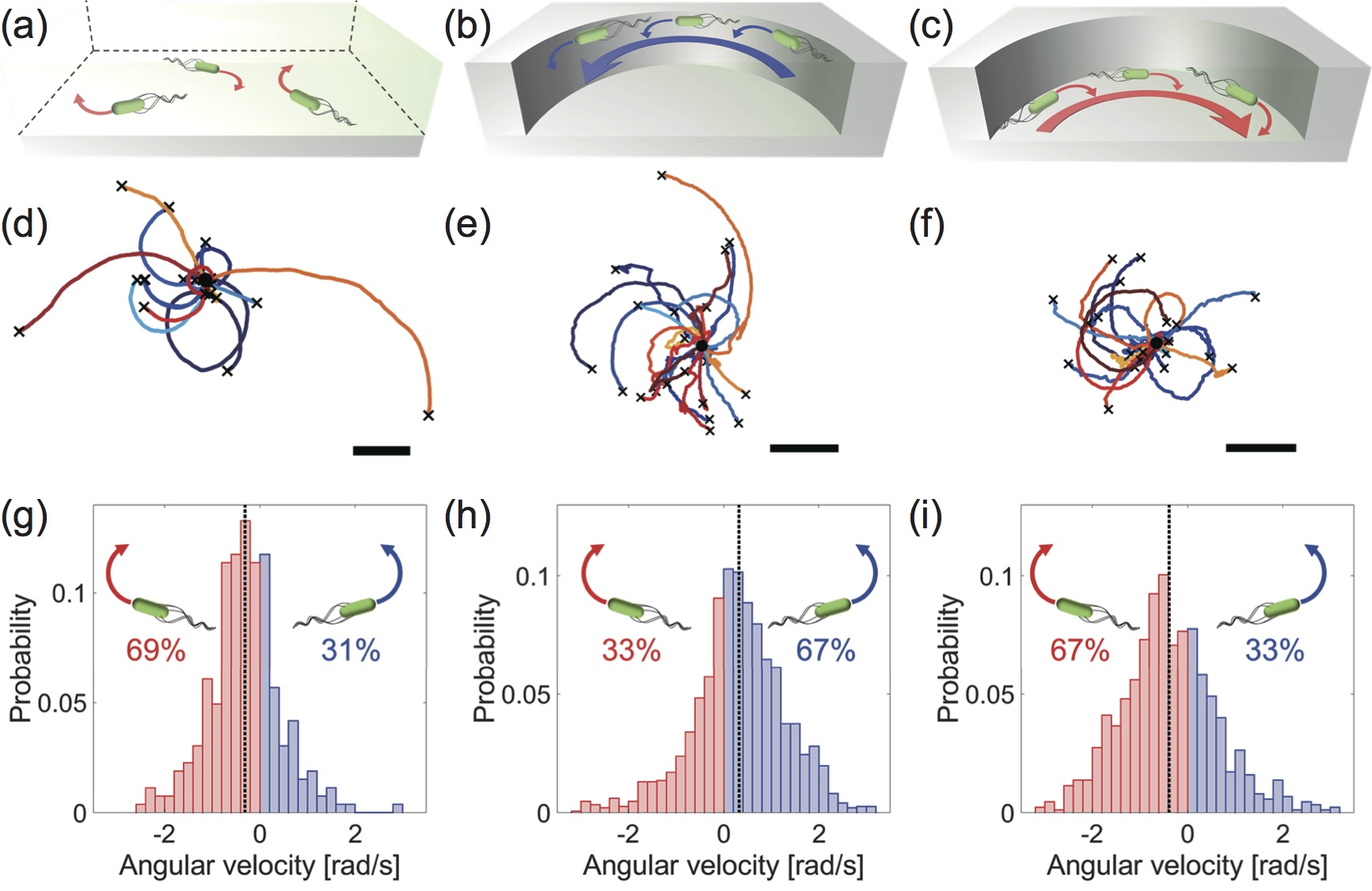}
\end{center}
\caption{\textbf{Intrinsic bacterial chiral swimming near interfaces.} (a-c) Schematic illustrations of (a) CW rotation (top view) of bacterial swimming on a solid interface with open lateral boundaries, (b) CCW rotation of bacterial swimming below oil/water interface, with lateral solid PDMS wall, and (c) CW rotation of bacterial swimming on the solid PDMS interface, with lateral solid PDMS wall. Scale bar in (a) is \SI{50}{\micro\meter}, and scale bars in (b) and (c) are both \SI{20}{\micro\meter}. (d-f) Representative trajectories of individual bacterial swimming in corresponding conditions ((d): open lateral boundaries, (e) top oil/water interface, (f) bottom solid PDMS interface), in dilute bacterial suspensions. The start points of each trajectory are fixed at the center of the plot, and their end points are indicated by a black cross. Colors code for different trajectories. (g-i) Histogram of the mean angular velocities of the detected trajectories in the corresponding cases ((g): open lateral boundaries, (h) top oil/water interface, (i) bottom solid PDMS interface). Proportions of rotation directions are given in blue for CCW and red for CW.}\label{fig.s4}
\end{figure}

Bacteria of the RP4979 strain swim persistently straight in the bulk of their environment due to their inability to tumble. However, in the vicinity of a solid interface, the torque generated by flagella rotation induces a CW (while looking from above, the bacteria being on top of the interface) deviation of the bacterial swimming that eventually leads to a circular trajectory. RP 4979 bacteria, in a dilute suspension, swimming on top of PDMS interface in the absence of lateral confinement show, as predicted, circular trajectories with a CW rotation direction (Fig. \ref{fig.s4}(a)(d)(g)). Such circular trajectory reflects the hydrodynamic interaction between the chiral rotation of flagella and boundaries.

A bacterial suspension was confined in PDMS microwells and sealed with an oil/water interface, which means that two interfaces were present in interaction with bacterial swimming. We thus tracked the swimming of individual bacteria in a dilute suspension, near the top oil/water interface, and the bottom PDMS interface of circular microwells (Fig. 3 in main text). Fig. \ref{fig.s4}(b)(e)(h) show the trajectory of bacteria that swim below the top oil/water interface, with top view. Their trajectories are curved towards the CCW direction. In addition, Fig. \ref{fig.s4}(c)(f)(i) shows the swimming trajectories near the bottom solid PDMS interface. 

\begin{figure}[tb]
\begin{center}
\includegraphics[scale=0.8,bb=0 0 450 602]{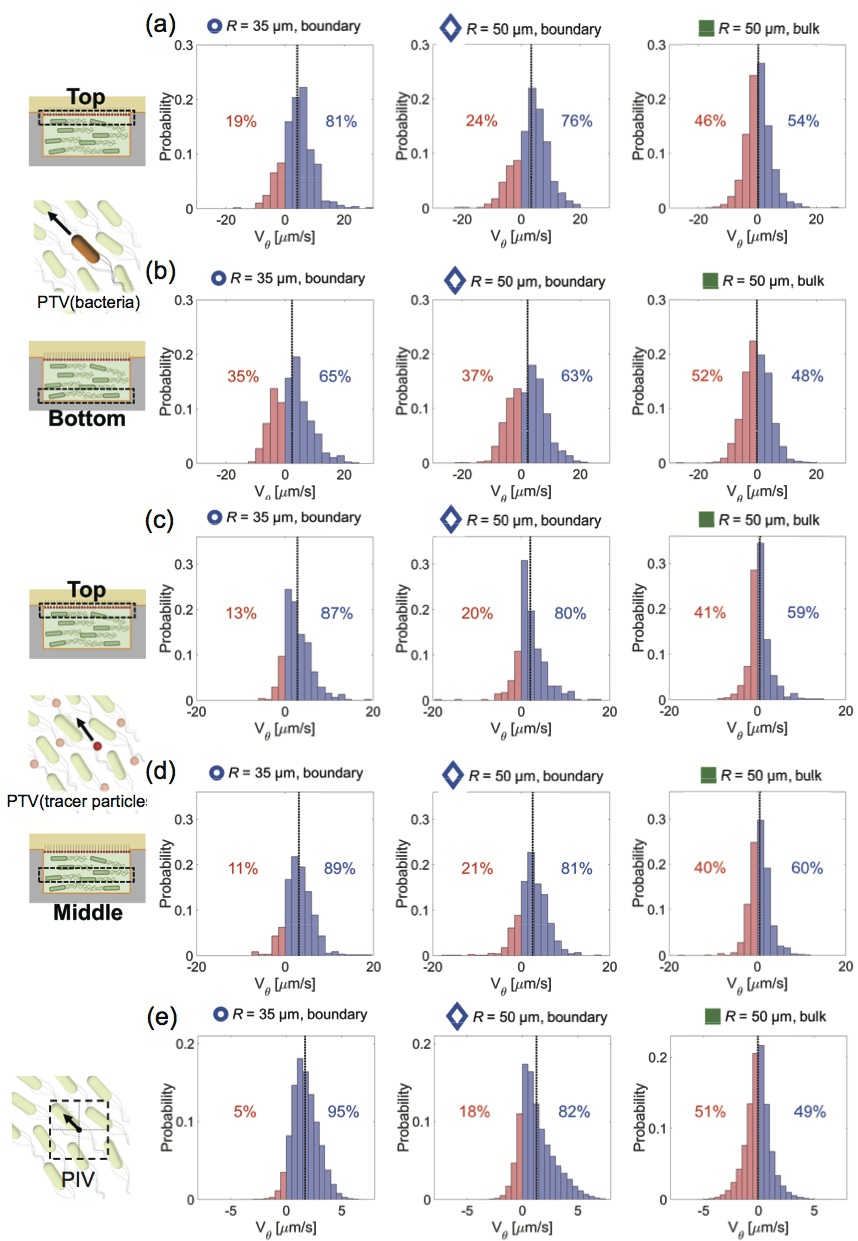}
\end{center}
\caption{\textbf{Histograms of azimuthal velocities in dense bacterial suspension.} (a and b) individual bacteria swimming near (a) the top oil/water interface and (b) the bottom PDMS interface, (c and d) fluid flow with tracer particles near (c) the top oil/water interface and (d) the middle of PDMS microwell, and (e) PIV velocity field of collective bacterial motion observed under bright field. The blue bars correspond to the probability of CCW rotation while the red bars to CW rotation. We also analyzed collective motion in two different sizes of microwells: $R = \SI{35}{\micro\meter}$ (small blue circle) and $R=\SI{50}{\micro\meter}$. In the larger microwells, we considered two regions: the boundary layer within \SI{10}{\micro\meter} from the lateral boundary (larger blue diamond), and the bulk that is more than \SI{10}{\micro\meter} away from the lateral wall (green filled square).}\label{fig.s5}
\end{figure}

Furthermore, we also analyzed the bacterial swimming, fluid flow, and collective motion (analyzed by PIV) at a dense ($20\% v/v$) bacterial suspension confined under asymmetric conditions.
Figs. \ref{fig.s5}(a) and (b) show the distribution of average azimuthal velocity $P(v_{\theta})$ at higher density. The small fraction of bacteria expressing dTomato (red fluorescent protein) was mixed at 1\% in dense bacterial suspension expressing YFP (yellow fluorescent protein) and the trajectory of individual bacteria was recorded by confocal microscopy. Fig. \ref{fig.s5}(a) (or  \ref{fig.s5}(b)) is the histogram of the azimuthal velocity of bacteria in top fluidic interface (or bottom solid interface). In Figs. 4(a) and 4(b) in main text, the chirality index $CI(v_{\theta}) = P(v_{\theta}) - P(-v_{\theta})$ is presented by using those data.  

Figs. \ref{fig.s5}(c) and (d) show the distribution of average azimuthal velocity $P(v_{\theta})$ of tracer particle (\SI{0.5}{\micro\meter}, red fluorescence) at higher density. The trajectories of particles were recorded at the top (\ref{fig.s5}(c)) or the middle (\ref{fig.s5}(d)) in the microwell by using confocal microscopy.

Finally, Figs. \ref{fig.s5}(e) is the distribution function of average azimuthal velocity $P(v_{\theta})$ from PIV analysis at higher density. One can find large fraction of bacteria (82-95\%) shows CCW rotation near boundary.

\begin{table}[htb] 
\begin{center}
\caption{Sample sizes and average velocities of histograms presented in Fig. \ref{fig.s4} (PTV (bacteria)).}
\begin{tabular}{|c|c|c|c|}
\hline
 & solid interface (Fig.\ref{fig.s4}(a)) & Top fluidic interface (Fig.\ref{fig.s4}(b)) & Bottom solid interface (Fig.\ref{fig.s4}(c))\\ \hline
spatial constraint & Boundary-free & Circular microwell & Circular microwell \\ \hline
sample size, N & N = 264 & N = 1469 & N = 876 \\ \hline
average angular velocity & \SI{-0.312}{\radian\per\second} & \SI{0.325}{\radian\per\second} & \SI{-0.385}{\radian\per\second} \\ \hline
\end{tabular}
\end{center}\label{table.s1}
\end{table}

\begin{table}[ht]
\begin{center}
\caption{Sample sizes $N$ and average velocities $\langle v_{\theta}\rangle$ of histograms presented in Fig. \ref{fig.s5}.}
\begin{tabular}{|c|c|c|c|ll}
\cline{1-4}
 & \textbf{$R$= \SI{35}{\micro\meter}, boundary} & \textbf{$R$= \SI{50}{\micro\meter}, boundary} & \textbf{$R$ = \SI{50}{\micro\meter}, bulk}  &  \\ \cline{1-4}
PTV (bacteria) TOP & $N = 491$, $\langle v_{\theta}\rangle = \SI{4.24}{\micro\meter}$/s  & $N = 784$, $\langle v_{\theta}\rangle = \SI{3.44}{\micro\meter}$/s & $N = 1129$, $\langle v_{\theta}\rangle = \SI{0.369}{\micro\meter}$/s &   \\ \cline{1-4}
PTV (bacteria) BOTTOM &  $N = 619$, $\langle v_{\theta}\rangle = \SI{2.35}{\micro\meter}$/s & $N = 1022$, $\langle v_{\theta}\rangle = \SI{2.16}{\micro\meter}$/s & $N = 1654$, $\langle v_{\theta}\rangle = \SI{-0.130}{\micro\meter}$/s &   \\ \cline{1-4}
PTV (tracer particles) TOP & $N = 442$, $\langle v_{\theta}\rangle = \SI{2.86}{\micro\meter}$/s & $N = 406$, $\langle v_{\theta}\rangle = \SI{2.06}{\micro\meter}$/s & $N = 628$, $\langle v_{\theta}\rangle = \SI{0.557}{\micro\meter}$/s &   \\ \cline{1-4}
PTV (tracer particles) MIDDLE & $N = 574$, $\langle v_{\theta}\rangle = \SI{3.12}{\micro\meter}$/s & $N = 736$, $\langle v_{\theta}\rangle = \SI{2.59}{\micro\meter}$/s  & $N = 1379$, $\langle v_{\theta}\rangle = \SI{0.541}{\micro\meter}$/s &   \\ \cline{1-4}
PIV & $N = 88536$, $\langle v_{\theta}\rangle = \SI{1.71}{\micro\meter}$/s & $N = 131240$, $\langle v_{\theta}\rangle = \SI{1.28}{\micro\meter}$/s & $N = 231200$, $\langle v_{\theta}\rangle = \SI{-0.048}{\micro\meter}$/s &   \\ \cline{1-4}
\end{tabular}\label{table.s2}\end{center}
\end{table}

\newpage

\subsection{The distribution of orientation angle in bacterial collective motion}
\begin{figure}[h]
\begin{center}
\includegraphics[scale=0.7,bb=0 0 268 285]{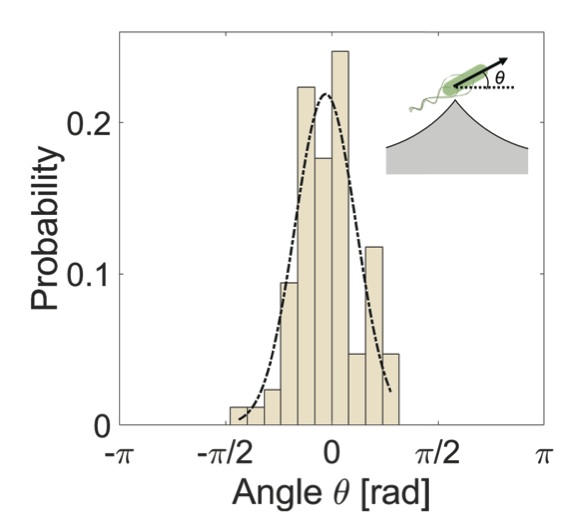}
\end{center}
\caption{\textbf{Orientation distribution of bacteria at high density.} A bacterial population was enclosed in a microwell with the boundary shape of two identical overlapping circles ($\Delta/R$ = 1.58). The group of bacteria exhibited collective motion in an FMV pattern, and we observed single bacteria labeled with fluorescent dTomato protein that swim away from the tip and the distribution of its heading angle was measured. The variance of this probability distribution is used to estimate the coefficient $\gamma_p$ of the polar alignment as $\sigma^2=\frac{D}{2\gamma \sin \Psi}$.}\label{figs_parameter}
\end{figure}

To analyze the fluctuation of the heading angle of bacteria $\theta(t)$ at a higher density, we measured the trajectory of single bacteria inside dense bacterial suspension confined in a microwell of overlapping two circles. The swimming bacteria show angular fluctuation $D$ as shown in Fig.\ref{figs_parameter}, but such fluctuation should be changed in the presence of orientation alignment due to collective motion. The strength of orientation interaction, which is defined as $\gamma_p$, is an important parameter for their collective motion. Indeed, the variance of the distribution function of heading angle $\sigma_{\theta}^2$ is closely related to both the coefficient of polar orientation interaction $\gamma_p$ and the fluctuation of bacterial heading angle $D$ as $\sigma_{\theta}^2=\frac{D}{2\gamma_p\sin\Psi}$ for FMV pattern ($\sigma_{\theta}^2=\frac{D}{2\gamma_p \cos\Psi}$ for AFMV pattern) because the polar orientation reorients the direction of bacterial swimming and in turn suppresses the angular fluctuation. The ratio between the angular fluctuation $D=0.12$ and the coefficient of polar alignment $\gamma_p$ is approximately equal to the variance of the distribution at the given geometric parameter $\Psi$. From the data for the straight swimming of bacteria in a bulk fluid (Fig.\ref{figs_parameter}), the angular diffusion coefficient $D$ is 0.12 rad$^2$/sec. In addition, the variance of angle distribution has 0.21. By using these values, we can estimate $\gamma_p$=$\frac{D}{2\sigma_{\theta}^2\sin\Psi}$=0.47. This value is used to examine the effect of chiral edge current for the pairing transition of FMV and AFMV, later. 

\newpage

\subsection{Active stirring of a micron-sized object in chiral bacterial vortex}

\begin{figure}[b]
\begin{center}
\includegraphics[scale=0.67,bb=0 0 545 483]{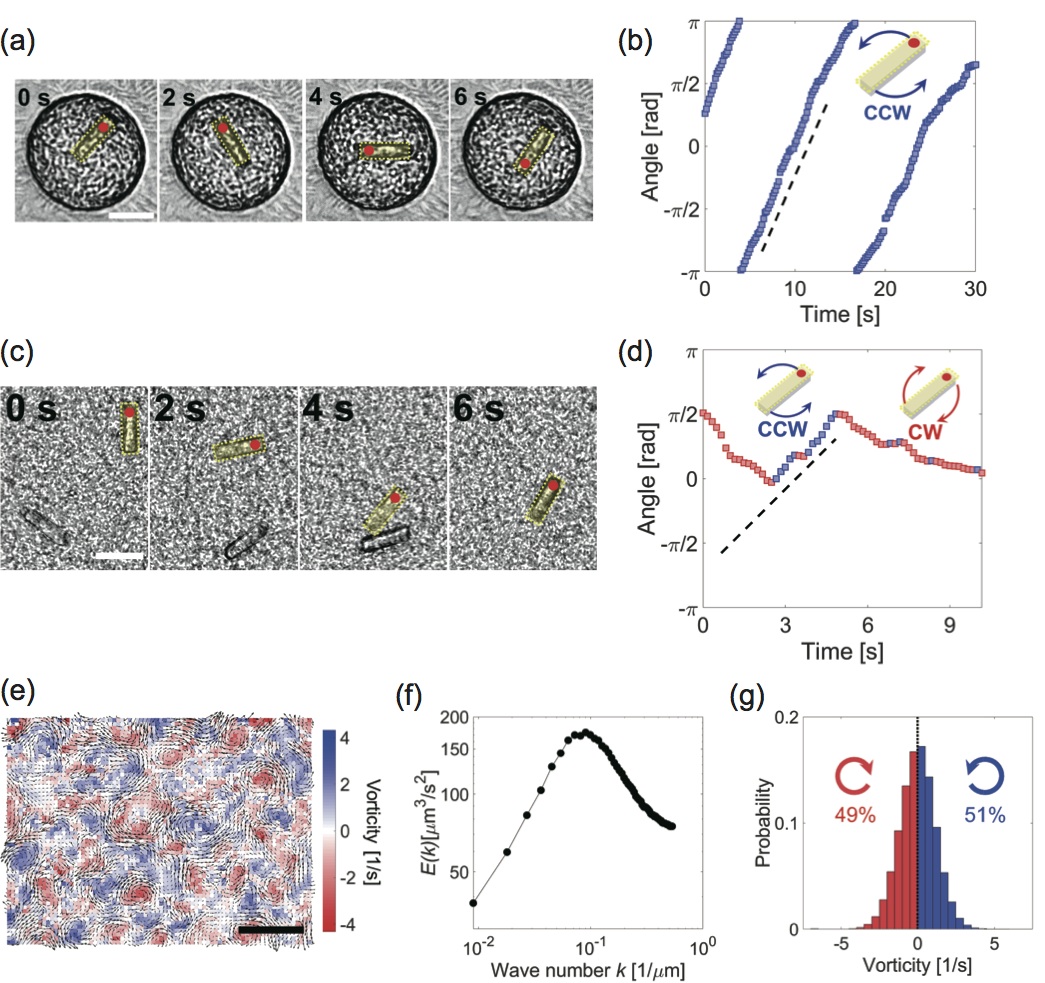}
\end{center}
\caption{\textbf{Rotation of rod-shaped object.} (a) Movement of a rigid rod in chiral bacterial vortex showed in (left) snapshots over $\SI{6}{\second}$. Red point indicates the head-end of the rod. Scale bar, \SI{20}{\micro\meter}. (b) Corresponding orientation angle increases linearly with time, exhibiting a constant CCW angular velocity (for comparison, dashed line has a slope of \SI{0.51}{\radian\per\second}). (c) Snapshots under bright field over \SI{6}{\second} of the random rotation and translation of a rigid rod in a dense bacterial suspension under quasi-two-dimensional confinement. Rod of interest is colored yellow in the image, and its head is arbitrarily defined by a red dot. Scale bar, \SI{20}{\micro\meter}. (d) Time evolution of the orientation angle of the marked rod. Blue color represents the CCW rotation and red color, the CW rotation. Both rotation directions have comparable angular velocities (black dashed line at \SI{0.67} {\radian\per\second}). Scale bar, \SI{20}{\micro\meter}. (e) Velocity field (black arrows) and vorticity map (red to blue colormap) obtained from the PIV analysis of the turbulent flow in a dense ($20\% v/v$) bacterial suspension under quasi-two-dimensional confinement between PDMS sheet (top) and a glass slide (bottom). Blue stands for CCW rotation and red stands for CW rotation. (f) Power spectrum of the previous velocity field. Peak is at \SI{0.09}{\per\micro\meter} which corresponds to a wavelength of \SI{35}{\micro\meter}. (g) Corresponding histogram of vorticity (Average vorticity \SI{0.00}{\per\second}). Proportions of rotation directions are given in blue for CCW and red for CW.}\label{fig.s3}
\end{figure}
Collective motion of suspended self-propelled particles enhances transport properties in active fluids, which has been used conjunctly with built-in geometry to direct the motion of objects larger than the suspended particles (e.g. gear-shape\cite{fabrizio,aronson} or ratchet-shape\cite{leonardo}). To further illustrate the transport properties chiral bacterial vortices, we confined a rigid rod (\SI{20}{\micro\meter} length and \SI{4}{\micro\meter} thickness, made of SU-8 photoresist) that is much longer than bacterial body (\SI{5}{\micro\meter} length and \SI{0.8}{\micro\meter} thickness). The rod consistently rotates in CCW direction over multiple rounds at \SI{0.5}{\radian\per\second}, i.e. 6 times faster than previously known ratcheted gears in a bacterial suspension (Fig. \ref{fig.s3}). As one would expect it, disordered vortices seen in quasi two-dimensional channel are also able to rotate rigid rods, larger than bacteria (Fig. \ref{fig.s3}(c)). However, the direction of rod rotation switches stochastically between CW and CCW and the absolute value of the angular velocity remains constant ($\simeq\SI{0.7}{\radian\per\second}$) (Fig. \ref{fig.s3}(d)). At high concentration, RP4979 strain bacteria present a turbulent behavior characterized by a large number of dynamic and intermingled vortices under quasi-two-dimensional confinement  (Fig. \ref{fig.s3}(e)). The PIV analysis of their collective motion reveals a widely distributed power spectrum with a peak at \SI{0.1}{\per\micro\meter} (Fig. \ref{fig.s3}(f)). This suggests that the vortices observed in this turbulent active flow have a size of typically \SI{35}{\micro\meter} or more. The PIV analysis of their collective motion also shows that the distribution of vorticity is even (Fig. \ref{fig.s3}(f)), which indicates that CW and CCW behaviors are identical under symmetric and quasi-two dimensional confinement. This equiprobability of CW and CCW leads to the stochastic change of rod rotation in two-dimensional chamber (Fig. \ref{fig.s3}(d)). Therefore, chiral bacterial vortex in present setup offers simple and fast material transport, even without built-in chirality such as gear-shape.

\subsection{FMV pairing order in triplet of circular microwells}
\begin{figure}[hbt]
\begin{center}
\includegraphics[scale=0.7,bb=0 0 641 425]{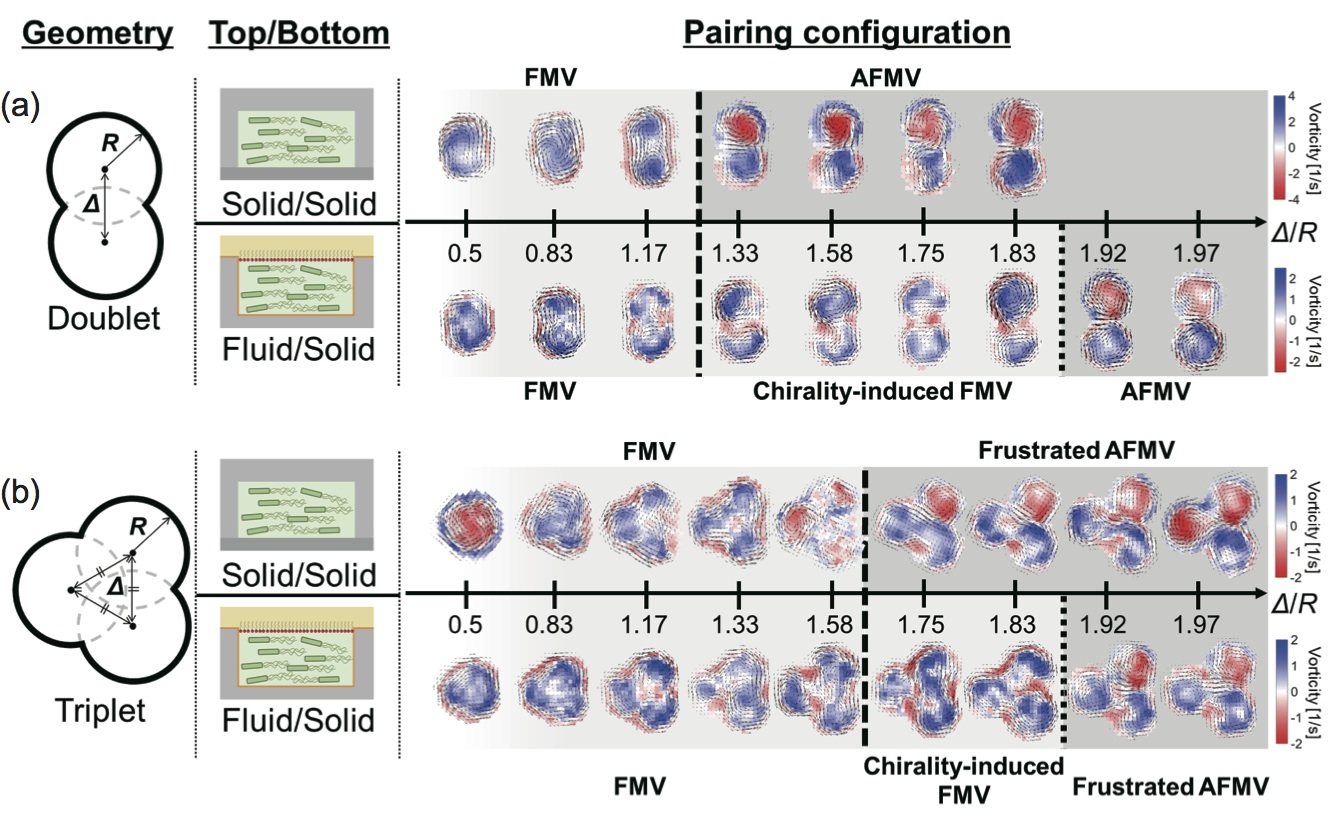}
\end{center}
\caption{\textbf{Edge current favors co-rotational vortex pairing}. Vortex pairing is affected by geometric frustration and chiral edge current. Dense bacterial suspension confined in multiple circular identical overlapping microwells ($R = \SI{19}{\micro\meter}$) can present various paring ((a) left and (b) left). In doublets of microwells ((a) top right and bottom right) no frustration is present and the effect of chirality shifts the transition point between FMV and AFMV pairing towards higher values of $\Delta/R$. In triplets of microwells (b), in the absence of chirality ((b) top right), FMV pairing is favored by geometric frustration and transits to AFMV pairing (frustrated order) at increased value of $\Delta/R$. When chirality is added to geometric frustration ((b) bottom right), FMV pairing is further favored, and the value of the transition point is further increased. The transition point from FMV to AFMV is $\Delta_c/R=1.9$ in both doublet and triplet circle microwells, which indicates that chiral edge current induces larger shift than the effect of frustration in triplet circle microwell.}\label{fig.s6}
\end{figure}

To determine the strength of the ordering induced by chiral bacterial vortices, we tested in against the rotational frustration observed in triplets of identical overlapping circular microwells. Here again, the microwell radius is noted $R$, and the center-to-center distance is noted $\Delta$ and is the same for all the pairs of microwells in the triplet. We used microwells with $R$ = \SI{19}{\micro\meter} and $0 \leq \Delta/R \leq 1.98$ (Figs. \ref{fig.s6}(a) and \ref{fig.s6}(b)).
When we confine a dense($20\% v/v$) bacterial suspension in such microwells, under symmetric conditions (top and bottom interfaces are in PDMS), geometric frustration is responsible for a shift of the FMV-AFMV (frustrated order) transition point $\Delta/R$ from 1.3-1.4 (observed in doublets of overlapping microwells under symmetric conditions, Fig. \ref{fig.s6}(a) top) to 1.6-1.7 (Fig. \ref{fig.s6}(b) top). Under asymmetric conditions (top interface is in oil/water and bottom interface is in PDMS), the chiral edge current of the interacting vortices further shifts that transition point to 1.8-1.9, similarly to what is observed in doublets of microwells (Figs. \ref{fig.s6}(a) bottom and \ref{fig.s6}(b) bottom). This result indicates that the chirality of bacterial vortices can generate larger unidirectional flow and overcome geometric frustration of interacting triplet vortices.

\newpage

\subsection{Order parameter of FMV pattern}
To analyze the ordered pattern of ferromagnetic vortex (FMV) pattern of bacteria, we used the order parameter $\Phi_{FMV}$ according to our previous study\cite{beppu}. Before giving the definition of $\Phi_{FMV}$, we need to derive analytical form of the angular velocity of interacting vortices inside a overlapping circular microwell. For this aim, we firstly find the analytical form for angular velocity of single vortex, ${v}_{\theta}(r)$ inside a circle of the radius $R$. We assume the boundary condition at $r=R$ is ${v}_{\theta}(R)$=0, and ${v}_{\theta}(r)$ is proportional to $r$. The spatial distribution of vorticity inside the circle is given by
\begin{equation}\label{angvel1}
  \omega(r) = \begin{cases}
    \omega & (0 \leq r \leq R-s) \\
    - \omega \Bigl[1- \frac{(R-s)^2}{R^2} \Bigr]  & (R-s \leq r \leq R)
  \end{cases}
\end{equation}
where $R-s$ is the position with maximum angular velocity and the size $s$ is \SI{4.6}{\micro\meter} estimated from experimental data. By solving \eqref{angvel1}, one can express the orthoradial velocity in a circular microwell $\bm{v}(r,\theta)$=$v_{\theta}(r)\bm{t}(\theta)$ as
\begin{equation}\label{angvel2}
\bm{v}(r,\theta) = \begin{cases}
    \frac{\omega}{2} \Bigl[1- \frac{(R-s)^2}{R^2} \Bigr]r \bm{t}(\theta)
 & (0 \leq r \leq R-s) \\
    \frac{\omega}{2} \Bigl(1- \frac{s}{R}\Bigr)^2\frac{R^2-r^2}{r} \bm{t}(\theta) & (R-s \leq r \leq R) \\
    0 & (r>R)
  \end{cases}
\end{equation}
where $\bm{t}(\theta)=(-\sin \theta, \cos \theta)$ is the unit orthoradial vector at the angular position $\theta$. In the following section, this analytic formulation is used to define the order parameter of FMV pattern.

Here we show the derivation of order parameter of ferromagnetic vortices (FMV) pattern used in Fig. 5(d). This order parameter reports the correlation of orientation field of velocity between the experimentally observed vortex pairing and numerically calculated FMV. For the numerical calculation of FMV pattern, the vortex confined in circular boundary is firstly considered: for each circle composing the doublet microwell, we set an index $j$, 1 stands for the left side and 2 for the right side. We define two sets of polar coordinates $(r_j, \theta_j)$; one for left circle is $(r_1, \theta_1)$ and the other for right circle is $(r_2, \theta_2)$. The origin of $j$ polar coordinates is set at the center of $j$ circle. We consider $\bm{t}_j(\theta_j)$ the orthoradial unit vector at the angular position $\theta_j$ centered on the center of the circle $j$ for $0\leq r_j \leq R$. In particular, we have $\bm{v}_j$$(r_j,\theta_j)=v_{\theta}(r_j)\bm{t}_j(\theta_j)$ where $v_{\theta}(r_j)$ is given by \eqref{angvel2}. 

\begin{figure}[th]
\begin{center}
\includegraphics[scale=0.7,bb=0 0 550 253]{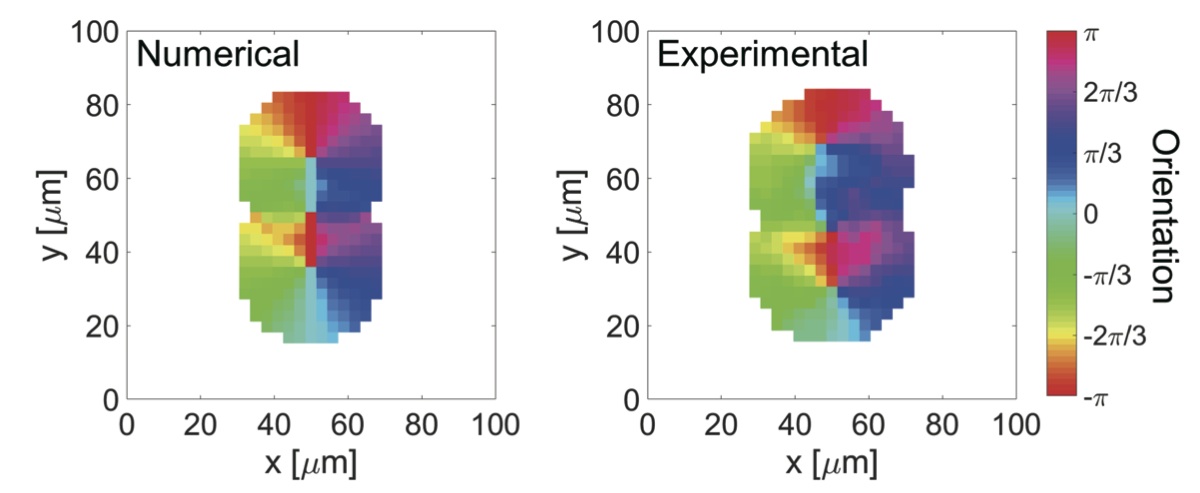}
\end{center}
\caption{\textbf{Orientation fields of FMV pattern.} (a) Orientation field of FMV pattern obtained from numerical method using Eq.\eqref{angvel4}. (b) Typical orientation field of interacting vortices in FMV pattern obtained in experiment.}\label{fig.Order}
\end{figure}

We then construct velocity field for vortices showing FMV pattern in the doublet microwell. In addition to the boundary condition of a doublet circle that is characterized by $R$ and $\Delta$, the polar coordinates $(r, \theta)$ defines the internal space. The origin of polar coordinates is placed at the centroid of the doublet shape and the velocity field, $\bm{v}$$(r, \theta)$, is in turn considered as the superposition of two vortices in $j$=1 and 2 circles. Because two vortices in FMV pattern show same angular velocities of $\bm{t}_1(\theta)=\bm{t}_2(\theta)$, we can describe the velocity field as
\begin{equation}\label{angvel3}
\bm{v}(r,\theta) = \sum_{j} \bm{v}_j(r_j, \theta_j) =  \sum_{j} v_{\theta}(r_j) \bm{t}_j(\theta_j)  .
\end{equation}

The orientation field of an FMV pattern lies on the unit vector $\bm{u}(r,\theta)$ such that 
\begin{equation}\label{angvel4}
\bm{u}(r,\theta) =\frac{\bm{v}(r,\theta)}{|\bm{v}(r,\theta)|}   .
\end{equation}

By using the inner product of expected orientation map $\bm{u}(r,\theta)$ and the one measured experimentally $\bm{p}(r,\theta)$, the order parameter $\Phi_{FMV}$ is then defined as
\begin{equation}
\Phi_{FMV}=\vert \langle \bm{p}(r,\theta)\cdot \bm{u}(r,\theta) \rangle \vert
\end{equation}
where $\langle \cdot \rangle$ denotes the ensemble average over all sites in a doublet microwell. One can estimate the deviation of experimentally obtained velocity field from ideal FMV pattern because $\Phi_{FMV} = 1$ if it matches the FMV pattern, and $\Phi_{FMV} = 0$ if it becomes an AFMV pattern or a disordered turbulence.

\newpage

\section{Theoretical analysis}

\subsection{Geometric rule of bacterial vortex transition with no edge current (without chirality)}

\begin{figure}[b]
\begin{center}
\includegraphics[scale=0.6,bb=0 0 659 210]{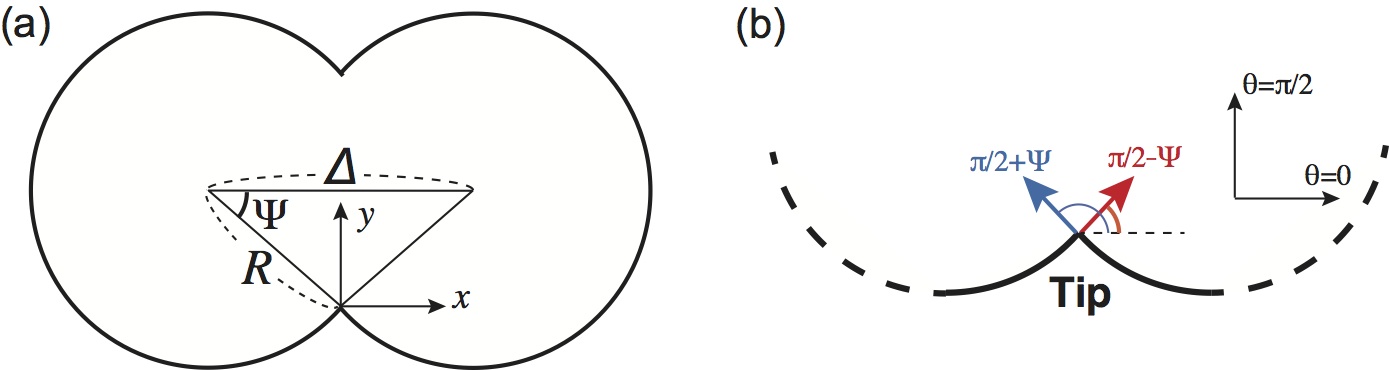}
\end{center}
\caption{\textbf{Boundary condition in theoretical model.} (a) The shape of the boundary conditions used in the theoretical model. Geometric parameters are shown in the figure. (b) The direction of movement and flow of particles along the wall near the tip. The horizontal right direction is defined as angle $\theta=0$, and the counterclockwise direction is a positive angle direction. The counterclockwise tangential direction of the left circular microwell (Red arrow) is an angle $\pi/2-\Psi$, and the clockwise tangential angle of the right circular microwell (Blue arrow) is $\pi/2+\Psi$.}\label{fig.s7}
\end{figure}

To explain the transition of ferromagnetic vortex (FMV) pattern and anti-ferromagnetic vortex (AFMV) patterns under geometric constraints, we construct theoretical model of orientational dynamics by considering the motion of self-propelled particles in a fluid. Bacteria is considered as a self-propelled point particle, with a position $\bm{r}_m(t) = (x_m(t), y_m(t))$ and an orientation $\bm{d}_m(t)$. For two-dimensional coordinates, the orientation of the particles is also expressed with the unit vector along the long-axis of bacteria $\bm{d}_m = \bm{d}(\theta_{m}) = (\cos\theta_m, \sin\theta_m)$. We impose the confinement of a circular microwell on the fluid and bacteria as shown in Fig.\ref{fig.s7}(a). The bacteria swim along the boundary wall at low noise limit and their heading angle $\theta_m$ is parallel to the tangential direction of boundary (Fig.\ref{fig.s7}(b)). The geometry of boundary shape is a pair of overlapping circular microwells with geometric parameter $\Delta/R$. The  ``tip'' of this doublet is a point where two bacterial vortices intersect and bacteria from different circular parts of the microwell collide. The parameters $\Delta$ and $R$ are also rewritten with the elevation angle $\Psi$ as $\cos\Psi = \frac{\Delta}{2R}$ .

The bacteria swim in their surrounding fluid and, when their density increases, their mutual interaction increases the alignment of their orientations. The evolution of $\bm{r}_m(t)$ is given by:
\begin{equation}
\dot{\bm{r}}_m(t) = v_0 \bm{d}(\theta_m)
\end{equation}
where particles move at a constant speed $v_0$.

We next consider that the orientation angle of the bacteria $\theta(t)$ is determined by to three distinct effects: mutual alignment of bacteria due to collision, hydrodynamic processes and random rotational diffusion. Swimming bacteria exert a force on the fluid and in turn induce fluid advection $\bm{v}(\bm{r},t)$$ \propto \bm{p}$, where $\bm{p}$ is a local polar vector defined by
\begin{equation}\label{polar}
\bm{p}(\bm{r}) = \langle\bm{d}(\theta_m)\rangle_{\bm{r}}.
\end{equation}
Collective motion occurs at a higher density and it generates the active fluid flow as $\bm{v}=V_0 \bm{p}$, with $V_0<v_0$. Such hydrodynamic flow rotates bacteria through velocity gradient effect.

The orientational dynamics of local alignment and rotation of bacteria by flow is 
\begin{equation}\label{orientation_all}
\dot{\theta}_m = - \gamma_p \underbrace{\sum_{\vert \bm{r}_{mn} \vert < \epsilon} \sin(\theta_m - \theta_n)}_{\rm polar \: alignment} - \gamma_n \underbrace{\sum_{\vert \bm{r}_{mn} \vert < \epsilon} \sin2(\theta_m - \theta_n)}_{\rm nematic \: alignment} + \gamma_a \underbrace{[\bm{d}_m \times (\bm{d}_m \cdot \nabla )\bm{v}]\cdot \bm{e}_z}_{\rm flow-induced \: alignment} +  \eta_m(t),
\end{equation}
where the first and the second terms govern the polar\cite{beppu}\cite{wioland1} and nematic\cite{Li} alignments of bacteria due to mutual collision, and the third term shows rotation by velocity gradient of fluid flow that generates the torque on bacteria\cite{hamby}\cite{Doi}. The forth term $\eta_m(t)$ represents the random fluctuation of the rotational direction, that satisfies $\langle \eta_m(t) \rangle$=0, $\langle \eta_m(t) \eta_n(t') \rangle$=$2D \delta_{mn}\delta(t-t')$ where $\delta_{mn}$ and $\delta(t)$ are Kronecker's delta symbol and Dirac's delta function, respectively. $D$ is the amplitude of the random noise that affects the orientation of bacteria.

As shown in Fig. 5 in main text, the geometrical feature of the boundary shape induces the transition from FMV to AFMV. However, it is not clear if this geometry-induced transition occurs due to the polar interaction (first term in Eq.(\ref{orientation_all})) or the nematic interaction (second term in Eq.(\ref{orientation_all})) depending on the orientation of the bacteria, or the flow-induced rotation change in fluid advection (third term in Eq.(\ref{orientation_all})). 

What we need to consider is the interaction of bacteria from left or right circles in a doublet microwell defined by geometric constant $\Psi$. In the tip where the two circles intersect, bacteria swimming in the left and right wells interact and become oriented. If bacteria swim in counterclockwise direction along the wall in the left microwell, their heading angle is $\pi/2 - \Psi$ at the tip, while bacteria swimming clockwise in the right microwell have an angle $\pi/2 + \Psi$ at the tip (Fig.\ref{fig.s8}(a)). 

In the following, a theoretical analysis is performed on how each orientational dynamics is affected by the geometric shape of the boundary. We assume the bacteria flowing from the left well to be aligned initially in CCW rotation, and hence the fluid flow follows the same rotation direction. In the case of AFMV (FMV) pattern, we consider the right well has CW (CCW) bacteria motion and flow rotation. 
To have the insight of the coupling, we assume it to be weak. In this way, we estimate the interaction between the vortices residing in different microwell by comparison with the swimming manner of bacteria and fluid flow of independent ones.
In addition, we assume the fluid flow of interacting vortices to be the linear superposition of the independent ones. In this way, we can consider how each term in Eq.~\eqref{orientation_all} affects the orientation of bacteria.
We first consider the effects of polar alignment due to mutual collision (including near field hydrodynamic interaction) and show that polar interaction can account for both (1) the preference of AFMV and FMV depending on geometrical parameter $\Psi$ and (2) the transition point consistent with experimental result. Because the effect of nematic alignment due to mutual collision cannot explain experimental results, the geometric dependence in the pairing order transition is decided by polar alignment. We then consider the flow-induced alignment of bacteria near the tip in order to explain the effect of edge current with rotational chirality. Finally, by extending the model of polar alignment with this chiral edge current, we propose geometric rule to account for both the selection of FMV pattern and the shift of the pairing order transition.

\subsubsection{The effect of polar alignment explains geometry-dependent FMV-AFMV transition}

At first, we consider the effect of polar alignment described by 

\begin{equation}\label{orientation_polar}
\dot{\theta}_m = - \gamma_p \sum_{\vert \bm{r}_{mn} \vert < \epsilon} \sin(\theta_m - \theta_n) +  \eta_m(t).
\end{equation}
where $\bm{r}_{mn}=\bm{x}_m-\bm{x}_n$ and $\epsilon$ is the effective radius of particle interaction. Polar interaction is involved in collective swimming in a highly dense suspension of bacteria, and a condition where the bacteria are oriented along the long axis with distinguishing the head and tail is favorable. Suppose the high-density of bacteria, the distribution of bacterial particles is homogeneous in space, and the density fluctuation is negligible. 

The aim of this theoretical analysis is to clarify whether polar interaction can explain the transition between FMV and AFMV patterns depending on the boundary shape $\Delta/R=2\cos\Psi$, and whether it also agrees with the transition point seen in experiment. Therefore, what we should consider is the collective motion under the spatial constraint where bacteria move along the walls of the left or right circles in a doublet microwell defined by geometric constant $\Psi$, and then collide mutually at the tip.

\begin{figure}[tb]
\begin{center}
\includegraphics[scale=0.6,bb=0 0 718 182]{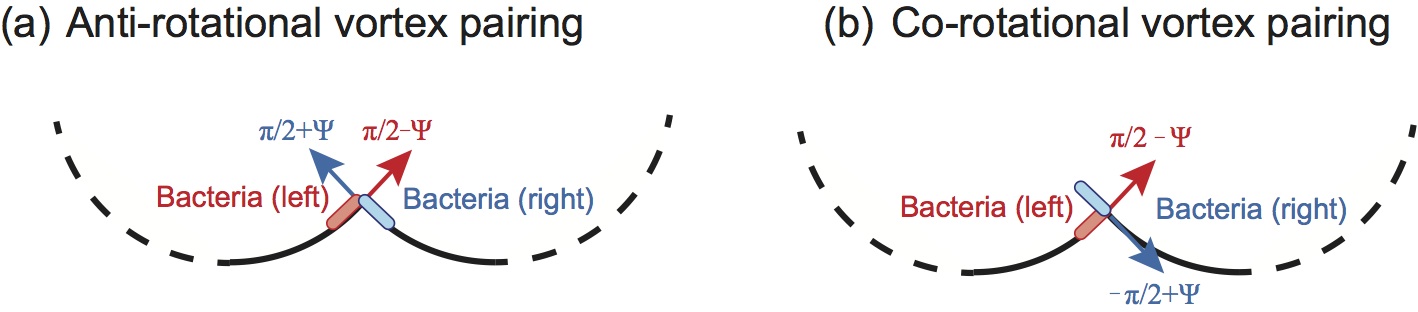}
\end{center}
\caption{\textbf{Polar alignment at the tip.} (a) Diagram of bacterial polar interaction showing anti-rotational vortex pairing at the tip. Bacteria moving along the left wall collide with bacteria moving from right circle at the tip. This pattern of polar interaction corresponds to AFMV pattern. (b) Diagram of bacterial polar interaction for co-rotational vortex pairing at the tip. The impact angle of the bacteria changes, and collective motion is in turn induced in the x-axis direction. This pattern eventually leads to FMV pattern.}\label{fig.s8}
\end{figure}

We first consider the AFMV pattern at which two bacterial vortices have opposite rotational direction. If bacteria swim in counterclockwise (CCW) along the wall in the left microwell, its heading angle is $\pi/2 - \Psi$ at the tip, while bacteria swimming clockwise (CW) in the right microwell have the angle $\pi/2 + \Psi$ at the tip (Fig.\ref{fig.s8}(b)). By taking mean field approximation\cite{beppu}\cite{vicsek}\cite{peruani}, the orientational dynamics Eq.\eqref{orientation_polar} is rewritten by 
\begin{equation}\label{orientation_polar2}
\dot{\theta}_m = - \gamma_p \bigl(\sin(\theta_m - \pi/2 + \Psi) + \sin (\theta_m - \pi/2 - \Psi)\bigr) +  \eta_m(t).
\end{equation}

%Figure S8 is showing polar interaction of bacterial particles at the tip

Then the dynamics of bacterial orientation in AFMV order is reduced to
\begin{equation}
\dot{\theta}_m(t)  = 2 \gamma_p \cos \theta_m \cos \Psi+ \eta_m(t).
\end{equation}
Once the orientational dynamics is obtained, one can derive the Fokker-Planck equation for the probability distribution of particle orientation, $P_{AFMV}(\theta,t;\Psi)$, in AFMV pattern as
\begin{equation}\label{polarFP_afmv}
\frac{\partial P_{AFMV}}{\partial t} = D \frac{\partial^2 P_{AFMV}}{\partial \theta^2} - 2\gamma_p \frac{\partial}{\partial \theta} \Bigl[(\cos\theta \cos \Psi) P_{AFMV} \Bigr].
\end{equation}

As for FMV pattern, the bacteria swimming in CCW rotation from the left circle has an angle $\pi/2 - \Psi$ at the tip while the bacteria entering the right circle has the heading angle of $- \pi/2 + \Psi$ at the tip (Fig.\ref{fig.s8}(b)). To construct the FMV pattern from Eq.\eqref{orientation_polar}, we take the same mean-field approximation and give the orientational dynamics of FMV pattern at the tip
\begin{equation}\label{orientation_polar3}
\dot{\theta}_m = - \gamma_p \bigl(\sin(\theta_m - \pi/2 + \Psi) + \sin (\theta_m + \pi/2 - \Psi)\bigr) +  \eta_m(t).
\end{equation}
Then the dynamics of bacterial orientation in FMV order is reduced to
\begin{equation}
\dot{\theta}_m(t)  = - 2 \gamma_p \sin \theta_m \sin \Psi+ \eta_m(t).
\end{equation}

The Fokker-Planck equation for the probability distribution of particle orientation, $P_{FMV}(\theta,t;\Psi)$, in FMV pattern is derived as
\begin{equation}\label{polarFP_fmv}
\frac{\partial P_{FMV}}{\partial t} = D \frac{\partial^2 P_{FMV}}{\partial \theta^2} + 2\gamma_p \frac{\partial}{\partial \theta} \Bigl[(\sin\theta \sin \Psi) P_{FMV} \Bigr],
\end{equation}

Because we focus on the static pattern of confined bacterial vortices, the left-hand side of both Eq.\eqref{polarFP_afmv} and Eq. \eqref{polarFP_fmv} is set at $\frac{\partial P}{\partial t} = 0$. By solving Eq.\eqref{polarFP_afmv} and Eq. \eqref{polarFP_fmv}, one can find the probability distribution $P_{AFMV}(\theta;\Psi)$ and $P_{FMV}(\theta;\Psi)$ at the steady state,
\begin{equation}\label{soln_afmv}
P_{AFMV}(\theta;\Psi)=\frac{\exp \bigl(\frac{2\gamma_p}{D}\sin \theta \cos \Psi \bigr)}{2 \pi I_0\bigl(\frac{\gamma_p}{D}\cos \Psi \bigr)}.
\end{equation}
and
\begin{equation}\label{soln_fmv}
P_{FMV}(\theta;\Psi)=\frac{\exp \bigl(\frac{2\gamma_p}{D}\cos \theta \sin \Psi \bigr)}{2 \pi I_0\bigl(\frac{\gamma_p}{D}\sin \Psi \bigr)}.
\end{equation}
where $I_0(\cdot)$ is first order Bessel function. In particular, AFMV pattern pointing in the vertical direction at $\theta = \pi/2 $ and the probability of FMV pattern pointing in the horizontal direction at $\theta = 0$ are both maximal. The unique solution $P_{FMV}(\theta = 0; \Psi_c) = P_{AFMV}(\theta = \pi/2; \Psi_c)$, meaning that AFMV and FMV patterns occur at equal probability,  is at the transition point $\Psi_c$. By comparing Eqs.\eqref{soln_afmv} and \eqref{soln_fmv}, the unique transition point is obtained as
\begin{equation}\label{geometricrule}
\sin \Psi_c = \cos \Psi_c,
\end{equation} 
which sets the transition at $\Psi_c = \pi/4$. Thus in the collective motion of self-propelled particles without chiral edge current, the geometric condition for the transition from FMV to AFMV is 
\begin{equation} \label{geometricrule2}
\Delta_c/R = 2 \cos \Psi_c = \sqrt{2}.
\end{equation}.
In addition, at $\Psi \leq \pi/4$, the probability of FMV pattern $P^{FMV}(\theta = 0; \Psi)$  is always larger than that of AFMV pattern $P^{AFMV}(\theta = \pi/2; \Psi)$, indicating that FMV pattern is favored at $\Delta/R\leq \sqrt{2}$. This geometric dependence is in excellent agreement with experimental result (Fig. 5(c) in main text and Fig.\ref{fig.s6}(a)), $\Delta_c/R \approx 1.3-1.4$.

This analysis suggests that polar alignment is considered to be the primary effect that can explain both the pattern formation of FMV and AFMV and its transition at $\Delta_c/R=1.3-1.4$ observed in the experiment.

$ $

\subsubsection{Nematic alignment cannot explain geometry dependent FMV-AFMV transition}
We next consider the effect of nematic alignment due to mutual collisions
\begin{equation}\label{orientation_nematic}
\dot{\theta}_m = -  \gamma_n \sum_{\vert \bm{r}_{mn} \vert < \epsilon} \sin2(\theta_m - \theta_n) +  \eta_m(t).
\end{equation}

We assume the left well has CCW chirality; bacteria swim in CCW direction, and fluid flows in CCW direction as well. Bacteria from left well are considered to have the orientation angle $\theta=\pi/2-\Psi$ at the tip. In the case of AFMV pattern, we assume bacteria from CW right well have the orientation angle $\theta=\pi/2+\Psi$
By taking mean field approximation of particle orientation, the dynamics of $\theta$ at time $t$ is given by
\begin{equation}\label{orientation_nematic-AFMV}
\dot{\theta}_m = - \frac{ \gamma_n }{2} \left\{\sin2(\theta_m - (\pi/2-\Psi)) + \sin2(\theta_m - (\pi/2+\Psi))\right\} +  \eta_m(t)=\cos 2\Psi \sin 2\theta_m + \eta_m(t)
\end{equation}
where $\gamma_n$ is the coefficient of nematic alignment. 

In the case of FMV pattern, CCW right well has the orientation angle $\theta=-\pi/2+\Psi$. In this case, the dynamics of $\theta$ at time $t$ is also given by
\begin{equation}\label{orientation_nematic-FMV}
\dot{\theta}_m = - \frac{\gamma_n}{2}  \left\{\sin2(\theta_m - (\pi/2-\Psi)) + \sin2(\theta_m - (-\pi/2+\Psi))\right\} +  \eta_m(t)=\cos 2\Psi \sin 2\theta_m +  \eta_m(t).
\end{equation}

Both AFMV and FMV follow the same dynamics of $\theta_m$ by comparing Eqns. \eqref{orientation_nematic-AFMV} and \eqref{orientation_nematic-FMV}, and one can find the interaction term is identical in $P_{AFMV}$ and $P_{FMV}$. FMV and AFMV patterns can be obtained with the same probability for all geometric parameters under nematic alignment of bacteria. In other words, the nematic interaction at the tip do not make any preference between AFMV and FMV. This analytical result is not consistent with the experimental results for geometrical transitions of FMV-AFMV patterns, suggesting that nematic orientation is not needed in orientational dynamics \eqref{orientation_all} for our experimental observation.

$ $

\subsubsection{Chiral edge current as advection-induced alignment}
\begin{figure}[b]
	\begin{center}
		\includegraphics[scale=1.4,bb=0 0 188 102]{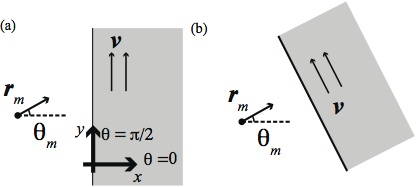}
	\end{center}
	\caption{\textbf{The alignment due to velocity gradient.} (a) A particle with orientation angle $\theta_m$ is affected by the velocity gradient of the fluid. Velocity of fluid is denoted by $\bm{v}=v_f \Theta(x)\bm{e}_y$. The calculation leads to the alignment of bacteria with the newly imposed flow direction, $\pi/2$. (b) When the imposed flow direction is given by $\theta_f$, bacteria is aligned with the direction of $\theta_f$.}\label{fig.fa}
\end{figure}

Bacteria experience torque due to velocity gradient, as given by
\begin{equation}\label{orientation_adv}
\dot{\theta}_m=\gamma_a [\bm{d}_m \times (\bm{d}_m \cdot \nabla )\bm{v}]\cdot \bm{e}_z+ \eta_m(t).
\end{equation}
Assuming step-like velocity field proportional to the step function $\Theta(x)$, we have $\bm{v}=v_f \Theta(x)\bm{e}_y$ as shown in Fig.\ref{fig.fa}. The bacteria heading to this step experiences torque, which can be calculated as
\begin{equation}\label{orientation_rot}
 [\bm{d}_m \times (\bm{d}_m \cdot \nabla )\bm{v}]\cdot \bm{e}_z=v_f\delta(x)(\cos^2 \theta)
\end{equation}
Bacteria whose orientation angle $\theta_m$ is heading to the step of velocity fields has normal velocity $v_0 \cos\theta$. Thus, the total change of the orientation $\Delta \theta$ due to the step-like flow fields can be obtained by
\begin{equation}\label{orientation_change}
\Delta \theta = \int dt \dot{\theta}= \frac{\gamma_a v_f }{v_0}\int_{-\infty}^{\infty} \cos \theta(x) dx = -\frac{\gamma_a v_f }{v_0} \sin\bigl(\theta_m-\frac{\pi}{2}\bigr).
\end{equation}
Thus, dynamics of the bacterial orientation under the effect of step-like flow field can be described as 
\begin{equation}\label{orientation-flow}
\dot{\theta}_m=-\gamma_e \sin\bigl(\theta_m-\frac{\pi}{2}\bigr)+ \eta_m(t),
\end{equation}
where $\gamma_e = \frac{ \gamma_a v_f }{v_0 \Delta t}$. When the direction of flow $\theta_f$ is arbitrary set to $\bm{v}=v_f (\cos \theta_f \bm{e}_x+\sin \theta_f \bm{e}_y)$, the same argument applies and the dynamics of the bacteria orientation is denoted as
\begin{equation}\label{orientation_change2}
\dot{\theta}_m=- \gamma_e \sin(\theta_m-\theta_f)+ \eta_m(t).
\end{equation}
where $\gamma_e$ is the coefficient for the alignment along the boundary edge ($\gamma_e\geq0$).

\begin{figure}[b]
	\begin{center}
		\includegraphics[scale=0.65,bb=0 0 718 182]{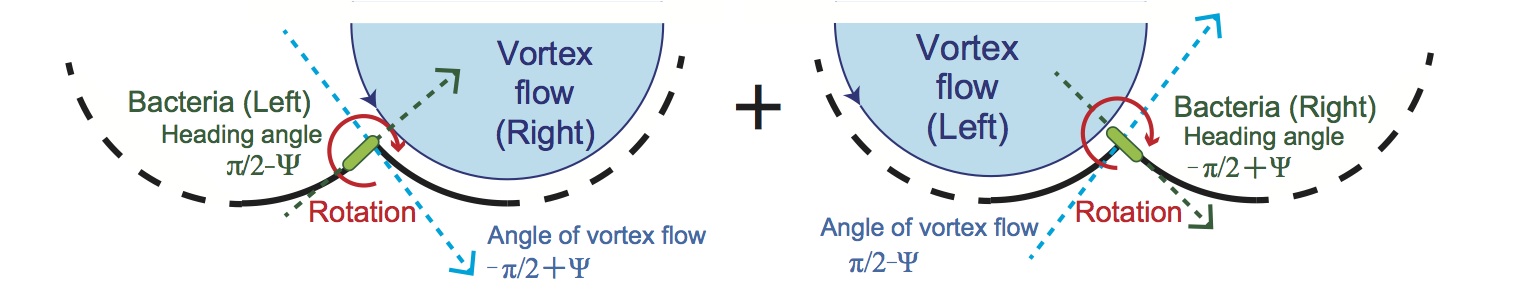}
	\end{center}
	\caption{\textbf{The edge current of bacteria and fluid streaming near boundary.} Diagram of interaction and superposition of bacteria and vortical flow explaining edge current in CCW direction. (left) Bacteria moving along the left wall collide with the vortical flow on the right microwell. (right) bacteria along the wall from the right collide with the vortical flow on the left microwell. Collective motion can be described with the sum of these two polar alignment of bacteria and vortical flow in co-rotational vortex pair. }\label{fig.s9}
\end{figure}

As we noted earlier, collective motion occurs at a higher density and it generates the active fluid flow along the polarized direction as $\bm{v}=V_0 \bm{p}$, and we assume the flow-like vortical rotation can be present at the steady state, with velocity $\bm{v}(\bm{r},t)$. At the tip, bacteria collide with the advection flowing at velocity $\bm{v}(\bm{r},t)$ as shown in Fig.\ref{fig.s7}. The direction of the flow advection is also directed to the tangential direction of the boundary wall, which is considered to be a stable vortical flow. 

In Figure 4 in main text, we found that chiral bacterial vortices show strong CCW bias in both bacterial swimming and fluid flow at boundary region. This fact allows one to propose the mathematical description of edge current of bacteria as follows: we consider the orientational dynamics under fluid advection for two interacting vortices with rotational flow in CCW direction. Since the bacterial population is constrained by a quasi-two dimensional space, the hydrodynamic flow is considered to be two-dimensional. The fluid flow in the microwell can be described as a superposition of the vortical flows present in the left and right circles of the doublet. In addition, we assume that bacteria do not collide with each other while they are aligned with the advective flow driven by collective motion that appears in a neighboring microwell. Then, it can be approximated that bacterial and flow alignment is limited to two combinations: the first is the alignment of bacteria moving in the left microwell along the vortical flow in the right microwell, and the second is the alignment case for bacteria moving in the right microwell with the vortical flow present in the left microwell (Fig. \ref{fig.s9}). 

As for edge current in CCW direction (Fig. \ref{fig.s7}(c)), the probability distribution functions $P_{L}(\theta,t;\Psi)$ and $P_{R}(\theta,t;\Psi)$ representing the angular distribution of bacteria in left and right are obtained, respectively (Fig. \ref{fig.s9}(a)). If the bacteria at the left side have the heading angle $\pi/2 - \Psi$ while the bacteria at the right side have the angle $-\pi/2 + \Psi$, the orientation probability distribution of bacteria can be obtained based on the equation describing orientational dynamics. 
The Fokker-Planck equations for probability of heading angle are
\begin{equation}\label{advFP_left_fmv}
\frac{\partial P_{L}}{\partial t} = D \frac{\partial^2 P_{L}}{\partial \theta^2} + \gamma_e \frac{\partial}{\partial \theta} \Bigl[\cos(\theta - \Psi)P_{L} \Bigr],
\end{equation}
and
\begin{equation}\label{advFP_right_fmv}
\frac{\partial P_{R}}{\partial t} = D \frac{\partial^2 P_{R}}{\partial \theta^2} + \gamma_e \frac{\partial}{\partial \theta} \Bigl[\cos(\theta + \Psi)P_{R} \Bigr].
\end{equation}

At this time, the bacteria swimming from the left microwell have an angle $\pi/2 - \Psi$, and the orientation changes due to the vortical flow of the right microwell. The vortical flow at the tip is directed along $- \pi/2 + \Psi$. Thereby, the change in orientation is given by $- \gamma_e \cos (\theta - \Psi)$. On the other hand, the bacteria entering the right microwell have an angle $- \pi/2 + \Psi$ at the tip, but in this case, the orientation is changed by being released from the flow at the angle $\pi/2 - \Psi$. Thus, the change in orientation is represented by $-  \gamma_e \cos (\theta + \Psi)$. 
Instead of finding the general solutions of Eqs. (\ref{advFP_left_fmv}) and (\ref{advFP_right_fmv}), we focus on the symmetry of probability distribution: because the shape of the boundary condition is highly symmetric at the tip, one can expect only negligible difference between $P_L(\theta,t)$ and $P_R(\theta,t)$ at this point. This geometric argument allows one to define the probability distribution of orientation for edge current by $P_{edge}=(P_{L}+P_{R})/2$. Hence, Eqs. \eqref{advFP_left_fmv} and \eqref{advFP_right_fmv} can be reduced to 

\begin{eqnarray}\label{advFP_fmv}
\frac{\partial P_{edge}}{\partial t} &=& D \frac{\partial^2 P_{edge}}{\partial \theta^2} +  \frac{\partial}{\partial \theta} \Bigl(\gamma_e \Bigl[\cos(\theta + \Psi)+\cos(\theta - \Psi)\Bigr]P_{edge}\Bigr) \nonumber \\
&=& D \frac{\partial^2 P_{edge}}{\partial \theta^2} + \frac{\partial}{\partial \theta} \Bigl(\Bigl[2\gamma_e \cos\theta \cos\Psi \Bigr]P_{edge}\Bigr),
\end{eqnarray}
where $P_{edge}(\theta,t;\Psi)$ is the probability distribution of particles at the tip with a heading $\theta$ at the time $t$ under CCW edge current. 

Eq.(\ref{advFP_fmv}) tells that the reorientation by the edge current has geometric dependence $\cos\Psi$. In addition, it tends to trap bacterial particle nearby boundary and rotate bacteria in CW direction (due to negative sign in right hand side). Due to such reorientation, bacteria keep swimming nearby boundary edge in CCW direction. In next section, in order to account for how chiral bacterial vortex favors FMV order and move the transition point, we consider the effect of particle reorientation due to this edge current in addition to polar alignment.

\subsection{Geometric rule of bacterial vortex transition with chiral edge current}
In this section, we theoretically explain that interacting chiral vortices favors FMV pattern in a wide range of geometric condition, by adding the effect of chiral edge current in CCW direction in the orientational dynamics. 
From the above analysis, one can propose the orientational dynamics of bacterial particle swimming close to boundary under the edge current in CCW direction and polar alignment at the tip position. Because the direction of chiral symmetry breaking originates from bacterial swimming onto surface, the direction of edge current should be continuously formed in the CCW direction regardless of whether the vortex pairing pattern is FMV or AFMV. When CCW-chiral collective motion remains near the wall, the heading angle tends to be downward, resulting in an edge current that continues along the wall as an FMV pattern. By taking mean-field approximation, the change of bacterial orientation under edge current with CCW rotation is described by $\dot{\theta} = -  \gamma_e ( \cos(\theta_m - \Psi) + \cos(\theta_m + \Psi)) = - 2 \gamma_e \cos\theta_m\cos\Psi$ that corresponds to the second term in right-hand side in Fokker-Planck equation Eq.\eqref{advFP_fmv}.
In addition, as shown in Eq.\eqref{orientation_polar3}, it is effective to make the polar interaction $\sum_{\vert \bm{r}_{mn} \vert < \epsilon} \sin(\theta_m - \theta_n) $ the primary effect in the orientational dynamics.  We thus start the orientational dynamics with edge current in CCW chirality by taking minimal Vicsek-style model :
\begin{equation}
\dot{\theta}_m = - \gamma_p \sum_{\vert \bm{r}_{mn} \vert < \epsilon} \sin(\theta_m - \theta_n) - 2 \gamma_e \cos\theta_m\cos\Psi + \eta_m(t).
\end{equation}

Then, one can write the Fokker-Planck equations for AFMV or FMV pattern with chiral edge current as follows:

for AFMV order with edge current:
\begin{eqnarray}
& &\frac{\partial P_{AFMV}}{\partial t}= D \frac{\partial^2 P_{AFMV}}{\partial \theta^2} - \frac{\partial}{\partial \theta} \Bigl(\bigl[2\gamma_p \cos \theta \cos \Psi -  2\gamma_e \cos \theta \cos \Psi \bigr] P_{AFMV}\Bigr),
\end{eqnarray}
and for FMV order with edge current:
\begin{eqnarray}
& &\frac{\partial P_{FMV}}{\partial t}= D \frac{\partial^2 P_{FMV}}{\partial \theta^2} + \frac{\partial}{\partial \theta} \Bigl(\bigl[2\gamma_p \sin \theta \sin \Psi + 2  \gamma_e \cos \theta \cos \Psi \bigr] P_{FMV}\Bigr).
\end{eqnarray}
As is apparent from these equations, the distribution function of the bacterial collective motion is affected by edge current and determined by the ratio of its rotational speed to the rotational diffusion constant $D$, and deviates the geometric rule of $\Delta_c/R = \sqrt{2}$ that characterizes the FMV-AFMV transition in the absence of chiral edge current.
The probability of realizing the FMV and AFMV patterns is obtained by
\begin{equation}
P_{AFMV}(\theta;\Psi)=A(\Psi) \exp \biggl[\frac{2}{D}((\gamma_p- \gamma_e)\sin\theta \cos\Psi)\biggr]. 
\end{equation}
and
\begin{equation}
P_{FMV}(\theta;\Psi)=B(\Psi) \exp \biggl[\frac{2}{D}(\gamma_p \cos\theta\sin\Psi + \gamma_e \sin\theta\cos\Psi)\biggr]
\end{equation}
where $A(\Psi)$ and $B(\Psi)$ are normalization factor with $\Psi$. Suppose that $\theta = 0$ is assigned to $P_{FMV}(\theta; \Psi) $, and $\theta = \pi/2$ is assigned to $P_{AFMV}(\theta; \Psi) $, we can rewrite
\begin{equation}
P_{AFMV}(\theta=\pi/2;\Psi)=A(\Psi)\exp\biggl[\frac{2}{D}\bigl((\gamma_p -  \gamma_e)\cos\Psi\bigr)\biggr]
\end{equation}
and
\begin{equation}
P_{FMV}(\theta=0;\Psi)=B(\Psi)\exp\biggl[\frac{2}{D}(\gamma_p \sin\Psi)\biggr].
\end{equation}

$A(\Psi)$ and $B(\Psi)$ are not equal unless $ \gamma_e = 0$, Then, the transition point with $P_{FMV}(\theta = 0; \Psi) = P_{AFMV}(\theta = \pi/2; \Psi)$ is given by
\begin{equation}\label{transition_chiral}
\log\biggl[\frac{A(\Psi)}{B(\Psi)}\biggr] + \frac{2(\gamma_p -\gamma_e)}{D} \cos\Psi =  \frac{2\gamma_p}{D}\sin\Psi .
\end{equation}

\begin{figure}[tb]
\begin{center}
\includegraphics[scale=0.5,bb=0 0 858 402]{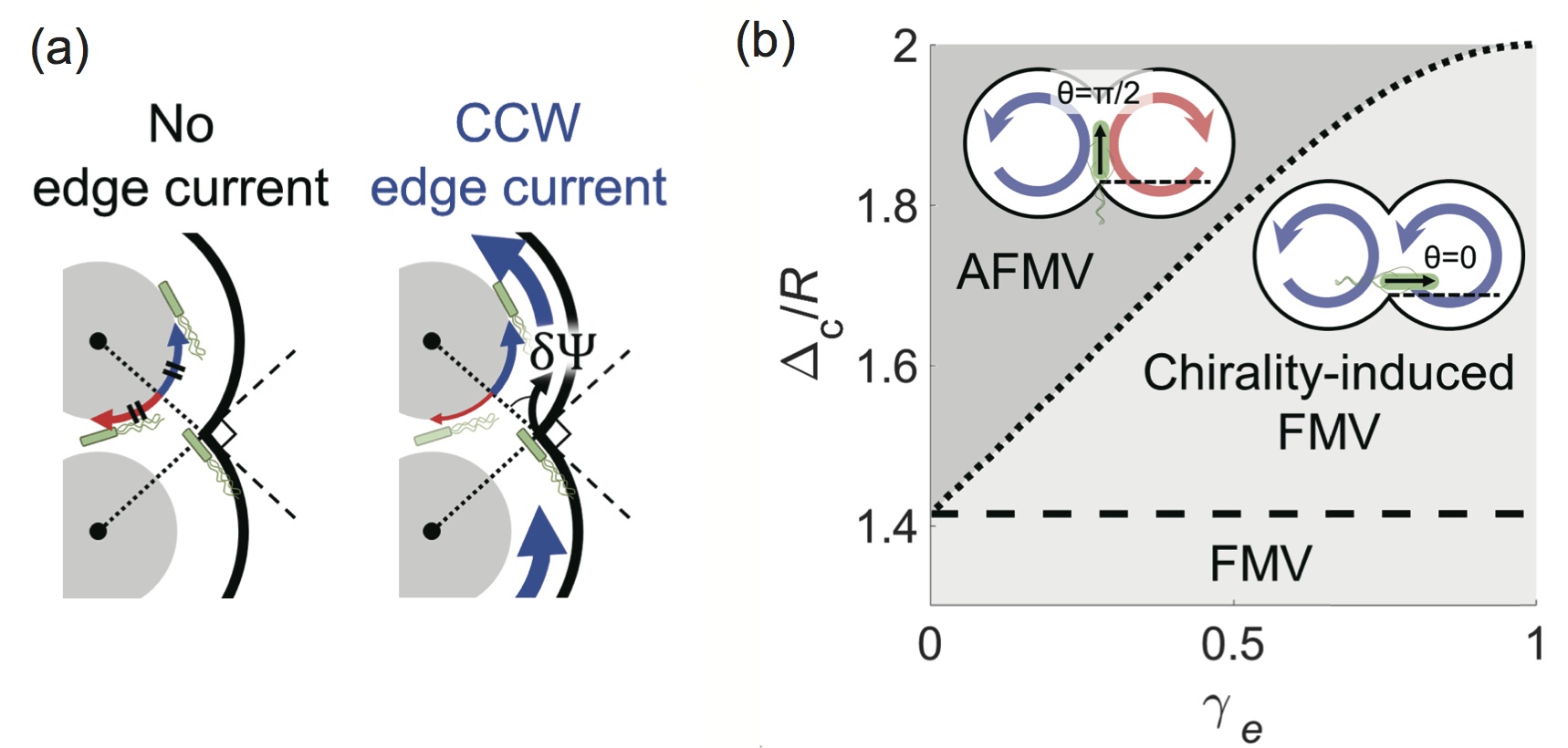}
\end{center}
\caption{\textbf{Phase diagram of vortex patterns.} (a) Illustration of the chiral bias of vortex pairing by edge current. In the configuration $\Psi =\pi/4$, that is $\Delta_c/R =2\cos(\pi/4)=\sqrt{2}$, CW and CCW rotations are equiprobable in both overlapping microwells in the absence of chirality effect (left). However, since the edge current creates a flow around the tip, the FMV pattern with CCW rotation is maintained, and the transition to the AFMV pattern shifts to the point of $\Delta_c/R\geq\sqrt{2}$ (right). (b) Phase diagram of FMV, chirality-induced FMV and AFMV is shown with geometric parameter $\Delta/R$ and the coefficient of chiral edge current $\gamma_e$. The dotted line indicates the transition point of chiral bacterial vortices and the horizontal broken line is the original transition point $\Delta/R=\sqrt{2}$ without chiral edge current.}\label{fig.s10}
\end{figure}

By calculating $\Psi$ satisfying the Eq. (\ref{transition_chiral}) numerically, the transition point to AFMV pattern can be obtained. FIG. \ref{fig.s10} is the phase diagram showing the transition point $\Delta_c/R$ with the chiral edge current. If bacterial vortex does not have such edge current, the transition point is $\Delta_c/R=\sqrt{2}$. In other word, the chiral edge current in bacterial vortex shifts the transition from FMV to AFMV at a point deviating from $\sqrt{2}$. If the transition from FMV pattern occurs at $\Delta_c/R > \sqrt{2}$ at non-zero $\gamma_e$, that is classified as the chirality-induced FMV. 

Interestingly, one can find that the shift of transition point increases in proportion to the coefficient of chiral edge current $\gamma_e$ in the range where chirality is small. In order to understand why such a linear relation holds, Eq. (\ref{transition_chiral}) was solved and the approximate solution of the transition point from chiral-FMV to AFMV was determined. To analyze the shift of transition point, we suppose the geometric parameter $\Psi$ that is close to $\cos\Psi=\sin\Psi=1/\sqrt{2}$ because this condition allows one to get $A \simeq B$ and then the first term in Eq.\eqref{transition_chiral} can approximate $\log\bigl[\frac{A(\Psi)}{B(\Psi)}\bigr] \simeq 0$. By solving Eq.(\ref{transition_chiral}) with  $\sin^2\Psi + \cos^2\Psi =1$,  the transition point $\Delta_c/R$ is
\begin{equation}\label{chiral_rule}
\frac{\Delta_c}{R} = \frac{2}{\sqrt{1 + \bigl(1 - \frac{ \gamma_e }{\gamma_p}\bigr)^2}}
\end{equation}
where $\Delta_c/R$ is in the range of $0 \leq \Delta/R \leq 2$ by definition. When Eq. (\ref{chiral_rule}) is linearized for $ \gamma_e /\gamma_p\ll1$, the transition point is rewritten as
\begin{equation}\label{chiral_rule2}
\frac{\Delta_c}{R} \approx \sqrt{2} \Bigl(1 + \frac{ \gamma_e }{2\gamma_p} \Bigr).
\end{equation} 
The obtained geometric relation Eq.(\ref{chiral_rule2}) means that the term related to chirality is added as a linear sum to the original expression of $\Delta_c/R=\sqrt{2}$ obtained when there is no chiral edge current. The shift of transition point is determined by the ratio between the effect of polar alignment $\gamma_p$ and the effect of edge current $ \gamma_e $. In addition, the transition point of chirality-induced FMV to AFMV is always larger than $\sqrt{2}$, suggesting that the edge current extends FMV order to a broader range of geometric conditions, in agreement with our experiments.

\end{document}